\def\zabs{$z_{\rm abs}$}

\def\ha{H-$\alpha$}
\def\hb{H-$\beta$}
\def\hg{H-$\gamma$}
\def\hd{H-$\delta$}
\def\lya{Ly-$\alpha$}

\def\oii{[O~{\sc ii}]}
\def\oiii{[O~{\sc iii}]}
\def\nii{[N~{\sc ii}]}
\def\sii{[S~{\sc ii}]}
\def\neiii{[Ne~{\sc iii}]}

\def\hi{H~{\sc i}}
\def\hii{H~{\sc ii}}

\def\nhi{\mbox{$\sc N(\sc H~{\sc I})$}}
\def\lognhi{\mbox{$\log \sc N(\sc H~{\sc I})$}}

\def\caii{Ca~{\sc ii}}

\def\cii{C~{\sc ii}}
\def\civ{C~{\sc iv}}
\def\feii{Fe~{\sc ii}}

\def\mgi{Mg~{\sc i}}
\def\mgii{Mg~{\sc ii}}

\def\siii{Si~{\sc ii}}
\def\siiii{Si~{\sc iii}}
\def\siiv{Si~{\sc iv}}
\def\tiii{Ti~{\sc ii}}
\def\niii{Ni~{\sc ii}}
\def\alii{Al~{\sc ii}}
\def\aliii{Al~{\sc iii}}
\def\znii{Zn~{\sc ii}}
\def\crii{Cr~{\sc ii}}
\def\coii{Co~{\sc ii}}
\def\oi{O~{\sc i}}

\def\ni{N~{\sc i}}

\documentclass[useAMS,epsfig]{mn2e}

\usepackage{rotating}
\usepackage{float}
\title[Neutral and Ionised Phase Metallicities]{A SINFONI Integral Field Spectroscopy Survey for Galaxy Counterparts to Damped Lyman-$\alpha$ Systems - V. Neutral and Ionised Phase Metallicities\thanks{Based on observations collected during programme ESO 87.A-0414 at the European Southern Observatory with SINFONI and X-Shooter on the 8.2 m telescopes operated at the Paranal Observatory, Chile.} }
\author[C\'eline P\'eroux et al.] {C\'eline P\'eroux$^1$\thanks{e-mail:celine.peroux@gmail.com}, 
Varsha P. Kulkarni$^2$ \& Donald G. York$^3$\\
$^1$ Aix Marseille Universit\'e, CNRS, LAM (Laboratoire d'Astrophysique de Marseille) UMR 7326, 13388, Marseille, France.  \\
$^2$ Dept. of Physics and Astronomy, Univ. of South Carolina, Columbia, SC 29208, USA.\\
$^3$ Dept. of Astronomy and Astrophysics and The Enrico Fermi Institute, University of Chicago, 5640 S. Ellis Ave, Chicago, IL 60637, USA.\\
}
\begin{document}

\date{Accepted 2013 October 24. Received 2013 October 23; in original form 2012 October 26}

\pagerange{\pageref{firstpage}--\pageref{lastpage}} \pubyear{2002}

\maketitle

\label{firstpage}

\begin{abstract}
The gas-phase and stellar metallicities have proven to be important parameters to constrain the star formation history of galaxies. However, \hii\ regions associated with recent star-formation may not have abundances typical for the galaxy as a whole and it is believed that the bulk of the metals may be contained in the neutral gas. It is therefore important to directly probe the metal abundances in the neutral gas, which can be done by using absorption lines imprinted on a background quasar. Recently, we have presented studies of the stellar content of a small sample of such quasar absorbers with \hi\ column densities measured to be in the sub-Damped Lyman-$\alpha$ to Damped Lyman-$\alpha$ range. Here, we present observations covering 300 nm to 2.5 $\mu$m of emission line spectra of three of these absorbing-galaxies using the long-slit spectrograph X-Shooter on the VLT. This allows us to compare the neutral and ionised phase metallicities in the same objects and relates these measures to possible signature of low-metallicity gas accretion or outflows of gas enriched by star formation. Our results suggest that the abundances derived in absorption along the line-of-sight to background quasars are reliable measures of the overall galaxy metallicities. In addition to a comparison of abundances in different phases of the gas, a potential observational consequence of differences in fueling mechanisms for disc galaxies is the internal distribution of their chemical abundances. We present some evidence for small negative metallicity gradients in the three systems. The flat slopes are in line with the differences observed between the two phases of the gas. These results suggest that a comparison of the \hi\ and \hii\ metallicities is a robust indicator of abundance gradients in high-redshift galaxies and do not favour the presence of infall of fresh gas in these objects.    

\end{abstract}

\section{Introduction}
%%%%%%%%%%%%%%%%%%%%%%%%%%%%%%%%%%%%%%%%%%%%%
%%%%%%%%%%%%%%%%%%%%%%%%%%%%%%%%%%%%%%%%%%%%%
\begin{table*}
\begin{center}
\caption{{\bf Journal of X-Shooter Observations.} The medium-resolution X-Shooter spectrograph covers the full wavelength range from 300 nm to 2.5 $\mu$m. Our observing strategy consisted in aligning the slit on both the bright background quasar and the faint absorbing-galaxy for which the exact sky positions is known from our previous SINFONI observations (P\'eroux et al. 2011a, 2012). The angular separation listed in the table is between the quasar and the absorbing-galaxy centre in arc sec and parsec.  }
\label{t:JoO}
\begin{tabular}{ccccccrlcc}
\hline\hline
Quasar 		   &Coordinates$^a$ &V Mag &z$_{\rm quasar}$&\zabs  &SFR &Angular &Separation &Observing Date &T$_{\rm exp}$ \\
&&&&& [M$_{\odot}$/yr] &["] &[kpc] &&[hrs]\\
\hline
Q0302$-$223 	         &03 04 49.86 $-$22 11 51.9	&16.0            &1.409	&1.0094      &1.8$\pm$0.6     &3.16 &25 &2011 Sep 8/9/11 &2.45 	\\	
Q0452$-$1640 	&04 52 13.60 $-$16 40 12.0	&18.0     	    &2.679	&1.0072      &3.5$\pm$1.0     &2.00&16	 &2011 Sep 9  &0.80 \\	
Q2352$-$0028 	&SDSSJ235253.51-002850.4 	&18.6            &1.624	&1.0318      &1.3$\pm$0.6     &1.50 & 12 &2011 Sep 8 &3.30  	\\	
\hline\hline 				       			 	 
\end{tabular}			       			 	 
\end{center}			       			 	 
\begin{minipage}{140mm}
{\bf Note:} \\
{\bf $^a$}SIMBAD coordinates except for the quasar which is part of SDSS, in which case the SDSS name is provided.
\end{minipage}
\end{table*}

{Estimating the amount of metals in galaxies and the cosmological evolution of this quantity is a key measure in the study of the formation of galaxies}. Metallicity is a fundamental property in the Universe: interstellar medium (ISM) elemental abundances and dust content; stellar populations; the evolution of galaxies and the intergalactic medium are all strongly linked to their metal content, as it controls the cooling function and the formation of the first stars and galaxies. Indeed, the chemical composition of galaxies is an indicator of both the history of star formation in them and the gas flows between the galaxies and the medium surrounding them, the so-called circumgalactic-medium (CGM). Observational measurements of the abundance of heavy elements in galaxies are therefore essential to our understanding of galaxy formation and evolution as well as the study of gas accretion and feedback to the galaxy's immediate environment. 

The abundances in the ISM are typically determined using emission-line spectroscopy of \hii\ regions. The primary indicator of chemical
evolution is typically traced by oxygen as its relative abundance surpasses all elements heavier than helium and it is present almost entirely in the gas phase (Snow \& Witt
1996; Savage \& Sembach 1996). Estimates of the metallicities of high-redshift galaxies are notoriously difficult because the indicators often used may be degenerate, and because different diagnostics are available at different redshifts, depending on the wavelength range covered (Kewley \& Ellison 2008). Since \hii\ regions are associated with recent star-formation, they may not have abundances typical for the galaxy as a whole. This is true in particular for star-forming galaxies, in which the bulk of the metals may be contained in the neutral gas. It is therefore important to directly probe the metal abundances in the neutral gas. This can be done using absorption lines imprinted on a background quasar. Indeed, the abundance measurements of the neutral gas detected in absorption are more robust than those for the gas detected in emission. However the absorption line measurements represent only one sightline through the galaxy (e.g. P\'eroux et al. 2006; York et al. 2006; Meiring et al. 2009). A reliable means of determining the origins of the absorbing gas is to obtain both the host-galaxy and absorption-line metallicity. Emission-line metallicities for a handful of quasar absorbers have been determined and are found to have nearly solar metallicities (e.g. Chen, Kennicutt \& Rauch 2005; Tripp et al. 2005; Gharanfoli et al. 2007; Kacprzak et al. 2010b).

In addition to a comparison of the abundances of both phases of the gas, a potential observational consequence of differences in fueling mechanisms for disc galaxies is their chemical abundance distributions. In the thin disc of the Milky-Way and other local
spirals, there is a negative radial metallicity gradient
(e.g. Searle 1971; Shields 1974; McCall, Rybski \& Shields
1985; Ferguson, Gallagher \& Wyse 1998). In contrast, the
thick disc of the Milky Way displays no radial abundance
gradient (Gilmore, Wyse \& Jones1995). Thick discs are ubiquitous in spiral
galaxies, and typically contain 10--25\% of their baryonic
mass. Merging events, which are believed to play a substantial
role in galaxy evolution (eg. de Ravel et al. 2009; L\'opez-Sanjuan
et al. 2011), seem to be a key physical process in shaping the
metallicity gradients of interacting galaxies (Rupke, Kewley \& Barnes 2010).
Recent observations have suggested that galaxies involved in
merging events show lower nuclear metallicities due to the infall
of pristine gas into the nucleus (Epinat et al. 2012). Metallicity gradients have been observed in the stellar populations
of early-type galaxies as well. They appear to correlate
with macroscopic properties such as, for example, stellar mass
(eg. Spolaor et al. 2009).

If the majority of the gas accretion in high-redshift
galaxies is from the IGM along filaments
which deposit pristine material onto
the galaxy disc at radii of 10--20 kpc, then the inner discs
of galaxies should be enriched by star-formation and supernovae. On the contrary, the outer-disc continually diluted by
pristine gas will leave strong negative abundance gradients
(Dekel et al. 2009) which will flatten as the gas accretion
from the IGM becomes less efficient. With the advent of integral field units (IFU), we now have new tools to probe the gradient of metallicity inside high-redshift galaxies (Cresci et al. 2010; Queyrel et al. 2012; Jones et al. 2010; Yuan et al. 2011, P\'eroux et al. 2011a).

Following an observational strategy pioneered by Bouch\'e et al. (2007, 2012) using the near-IR 3D SINFONI spectrograph on the VLT, we have previously reported \ha\ and \nii\ emission lines for a small sample of five absorbing-galaxies associated with high \nhi-column density quasar absorbers (P\'eroux et al. 2011a, 2012). Here, we present emission-line spectra for three of these objects covering the observed wavelength 300 nm to 2.5 $\mu$m with long-slit spectroscopy of the X-Shooter on the VLT. The data allow to study the \hii\ regions of these z$\sim$1 galaxies and derive their abundances and dust content. We use absorption lines tracing the neutral gas at small impact parameter from previous high-resolution observations to probe for metallicity of the neutral gas in the same systems and the amount of depletion due to dust in this phase. This allows us to compare metallicities in the neutral and ionised phases in the same objects, to study the spatial distribution of metals and relate these measures to possible signatures of low-metallicity gas accretion or outflows of gas enriched by star formation.

The present paper is structured as follows. A summary of observational details and data reduction steps are provided in Section 2. 
In the third section, we first present the metallicity for each absorbing galaxy in both the neutral and the ionised phase, as well as put constrains on the AGN contamination in the sample and derive the ionisation parameter for these objects. We then proceed with an analysis of the internal gradients of the galaxies measured in the ionised phase and a comparison of the metallicity measured in emission and in absorption. Finally, the study of the dust content of the absorbers in both phases is presented in Section 4. Throughout this paper, we assume a cosmology with H$_0$=71 km/s/Mpc, $\Omega_M$=0.27 and $\Omega_{\rm \Lambda}$=0.73.

\begin{table*}
\begin{center}
\caption{{\bf Neutral gas-phase abundances from the literature for the three DLAs/sub-DLAs targeted in this work}.  These metallicities with respect to solar values are measured in absorption along the line-of-sight to the background quasar. 'Inst' refers to the spectrograph used to take these data: Keck/HIRES, VLT/UVES and Magellan/MIKE.}
\label{t:HI_Metals}
\begin{tabular}{cccccccccccc}
\hline\hline
Quasar 		&Inst  &\lognhi &[Zn/H] &[Si/H] &[Mn/H] &[Cr/H] &[Fe/H] &[X/H]$^{\dagger}$  &Ref \\
\hline
Q0302$-$223 &HIRES	&20.36$\pm$0.11 &$-$0.56$\pm$0.12 &$-$0.73$\pm$0.12 &$-$1.32$\pm$0.12 &$-$0.96$\pm$0.12 &$-$1.20$\pm$0.12 &$-$1.04$\pm$0.12 &(a)\\
Q0452$-$1640 &UVES	&20.98$^{+0.06}_{-0.07}$&$-$0.96$\pm$0.08 &$-$0.66$\pm$0.11 &$-$1.54$\pm$0.11 &$-$1.14$\pm$0.09 &$-$1.34$\pm$0.08 &$<-$2.11 &(b)\\
Q2352$-$0028 &MIKE	&19.81$^{+0.14}_{-0.11}$ &$<-$0.51 &$+$0.14$\pm$0.14&$<-$1.37 &$-$0.49$\pm$0.19 &$-$0.37$\pm$0.13 &$>-$0.40 &(c)\\
\hline\hline 				       			 	 
\end{tabular}			       			 	 
\end{center}			       			 	 
\begin{minipage}{170mm}
$^{\dagger}$: [X/H] where X is Ni, Ti and Mg for Q0302$-$223, Q0452$-$1640 and Q2352$-$0028, respectively.\\
{\bf References:} (a) Pettini et al. 2000; (b) P\'eroux et al. 2008 and (c) Meiring et al. 2009. 
\end{minipage}
\end{table*}

\section{Observations and Data Reduction}
%%%%%%%%%%%%%%%%%%%%%%%%%%%%%%%%%%%%%%%%%%%%%
%%%%%%%%%%%%%%%%%%%%%%%%%%%%%%%%%%%%%%%%%%%%%

We have observed with X-Shooter three of the five absorbing-galaxies (namely Q0302$-$223, Q0452$-$1640 and Q2352$-$0028) discovered with SINFONI in P\'eroux et al. (2011a, 2012). Data for the two additional systems towards Q1009$-$0026 and Q2222$-$0946 were requested but not taken and not granted. A journal of observations summarising the target properties and experimental set-up is presented in Table~\ref{t:JoO}. The table provides the exposure times for each of the objects. The observations were carried out in service mode (under programme ESO 87.A-0414) at the European Southern Observatory with X-Shooter (Vernet et al. 2011) on the 8.2 m KUEYEN telescope. The medium-resolution X-Shooter spectrograph covers the full wavelength range from 300 nm to 2.5 $\mu$m thanks to the simultaneous use of three spectroscopic arms (UVB, VIS and NIR). Our observing strategy consisted in aligning the slit on both the bright background quasar and the faint absorbing-galaxy for which the exact sky positions are known from our previous SINFONI observations. We used the long-slit mode with slit width of 1.0" for UVB and 0.9" for the VIS and NIR arms. With these settings, the expected spectral resolution is 59 km/s (UVB), 34 km/s (VIS) and 53 km/s (NIR) respectively. Our actual resolution is only slightly higher than thes estimates because the typical seeing (0.8") is only slightly smaller that the slit widths. To optimise the sky subtraction in the NIR, the nodding mode was used following an ABBA scheme with a nod throw of 5 arcsec. The total exposure time was divided in Observing Blocks of 2960 seconds each. One additional exposure towards Q2352$-$0028 was classified as ``executed'' (i.e. not within the observer requirements according to ESO's classification) but turns out to be useful for our analysis.

The data were reduced with the latest version of the ESO X-Shooter pipeline, version 1.5.0 (Goldoni et al. 2006) and additional external routines for the extraction of the 1D spectra and their combination. Master bias and flat images based on data taken closest in time to the science frames were used to correct each raw spectra. Bias and flat-field correction are part of the ESO pipeline. Cosmic rays were removed by applying the Laplacian edge technique of van Dokkum (2001) and sky emission lines are subtracted using the Kelson (2003) method. The orders for each arm are then extracted and rectified in wavelength space using a wavelength solution previously obtained from calibration frames. The 2D orders are merged using the errors as a weight in the overlapping regions. We found the 1D extraction from the ESO pipeline unsatisfactory and performed our own extractions using the 'apall' routine within IRAF and interactively optimising the signal window definition and background regions for subtraction. For each exposure, we extracted a quasar spectrum on one hand, and a spectrum of the absorbing-galaxy on the other hand (for the VIS and NIR arm only in the latter case). We then merged the 1D spectra weighting each spectrum by the signal-to-noise ratio.  

The spectra were flux-calibrated using a spectrophotometric standard star. In some cases, the UVB spectrum is scaled to the VIS spectrum to match in the overlapping regions around 500 nm. 
We did not apply slit corrections for losses due to a finite slit width. The resulting flux calibration is compared with the quasar 2MASS magnitudes in the NIR and V magnitudes in the VIS in order to estimate the flux uncertainties. In the case of Q2352$-$0028, we compare our flux calibration to the flux calibrated spectrum from the Sloan Digital Sky Survey (SDSS) and find that our calibration gives a flux lower by a factor of two compared to the Sloan calibration. This is not unexpected given the typical variation in flux of quasars and the 10 years elapsed between the SDSS spectrum and the X-Shooter spectrum. However, we note that the shapes of the two spectra match very well. In order to perform the analysis of the absorption lines associated with the galaxies, the quasar spectra were normalised using a spline function going through regions devoid of absorption features. For the absorbing-galaxy spectra, the \ha\ emission lines fluxes are compared with SINFONI's and found to be consistent within the errors.

We correct the wavelength calibration to a vacuum heliocentric reference. As in Noterdaeme et al. (2012), we note a constant 0.5 pixel shift in the VIS arm with respect to the UVB arm and with respect to higher-resolution spectroscopy from VLT/UVES, Keck/HIRES or Magellan/MIKE. This is not observed in the NIR arm, where the SINFONI data allow for a direct comparison. The VIS spectra are thus redshifted to match the other information available. The resulting agreement between all three sources of information (X-Shooter, high-resolution spectroscopy and NIR IFU SINFONI data) is excellent. We also note that having both the absorption features of the DLAs/sub-DLAs in the quasar spectrum and the emission lines from the absorbing-galaxy on the same wavelength solution allows for a robust comparison of the absorption/emission velocity shifts. 

%%%%%%%%%%%%%%%%%%%%%%%%
%%%%%%%%%%%%%%%%%%%%%%%%
\section{Metallicity}
\subsection{Integrated Neutral Phase Metallicity}

\begin{figure*}
\begin{center}
\includegraphics[height=5.5cm, width=9.5cm, angle=-90]{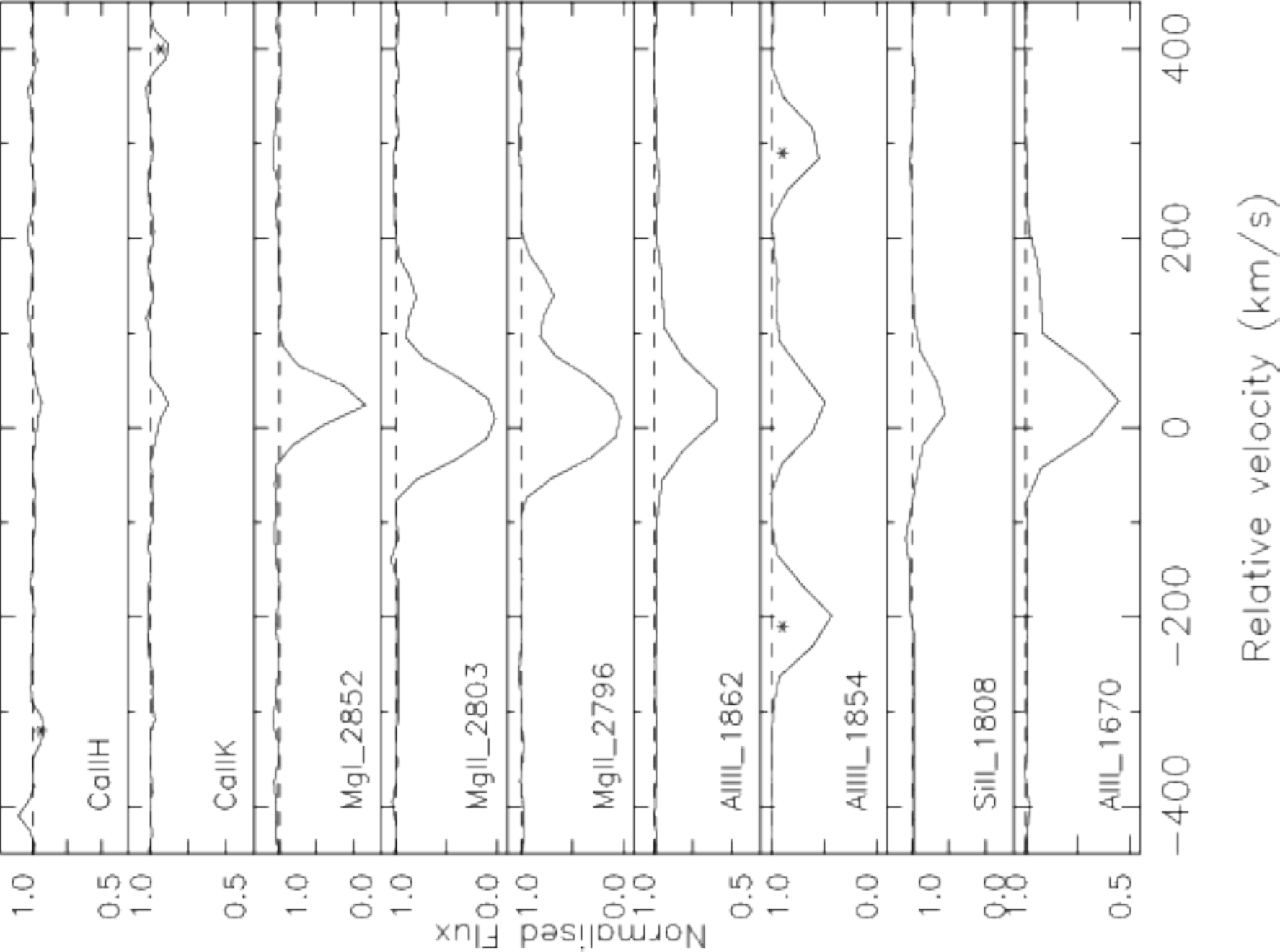}
\includegraphics[height=5.5cm, width=9.5cm, angle=-90]{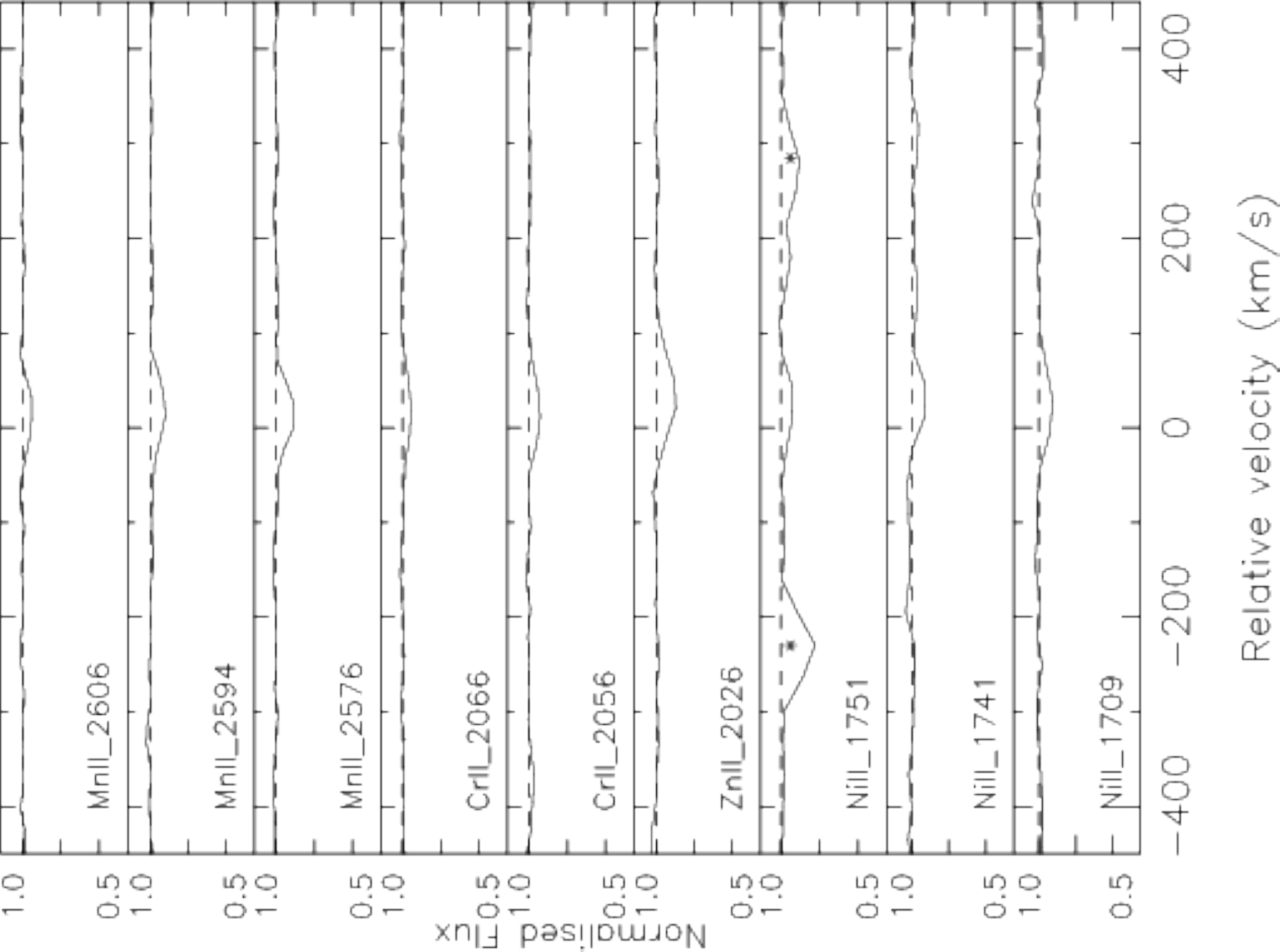}
\includegraphics[height=5.5cm, width=9.5cm, angle=-90]{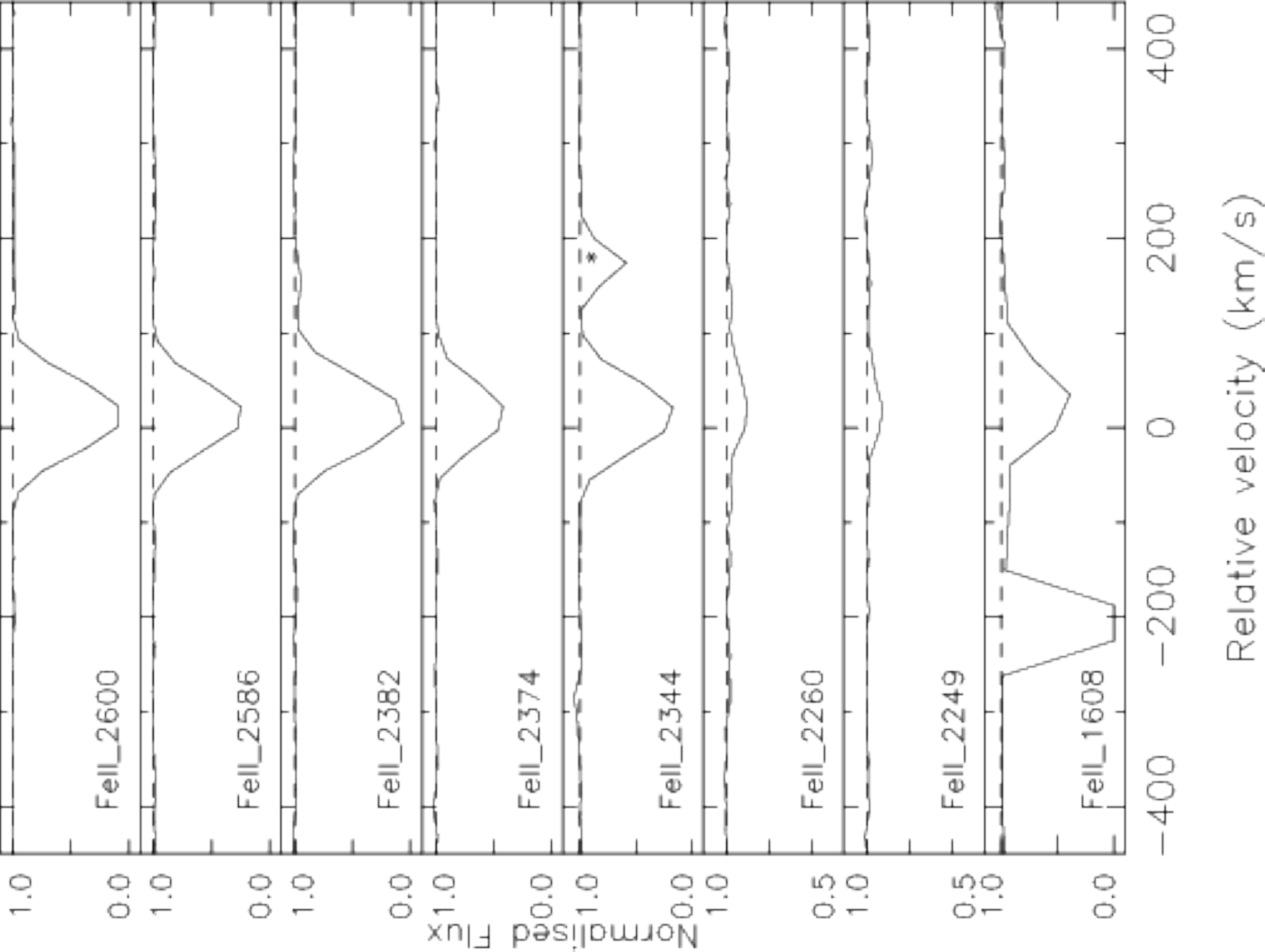}
\caption{{\bf Normalised X-Shooter 1D quasar spectrum of the absorption lines in the DLA  towards Q0302$-$223.} In this and the following figures, the y-axis has different scales in different panels. \feii\ lines from $\lambda$ 1608 to 2600 \AA\ are seen as well as all the many Fe-peak elements already reported by Pettini et al. (2000) at higher-resolution from Keck/HIRES spectroscopy. We report 3 additional weak lines of \niii\ ($\lambda \lambda \lambda$ 1709, 1741, 1751) as well as \aliii\ ($\lambda$ 1670) and \aliii\ ($\lambda \lambda$ 1854 and 1862). Finally, the \caii\ H and K doublet is detected in the NIR arm part of the spectrum. The features marked as '*' in panels \niii\ $\lambda$ 1754, \alii\ $\lambda$ 1854 and \feii\ $\lambda$ 2344 are features in the \lya\ forest lying near our target lines while the ones in the \caii\ H and K are due to tellurics. 
}
\label{f:Q0302_Abs}
\end{center}
\end{figure*}

\begin{figure}
\begin{center}
\includegraphics[height=8.5cm, width=16.cm, angle=-90]{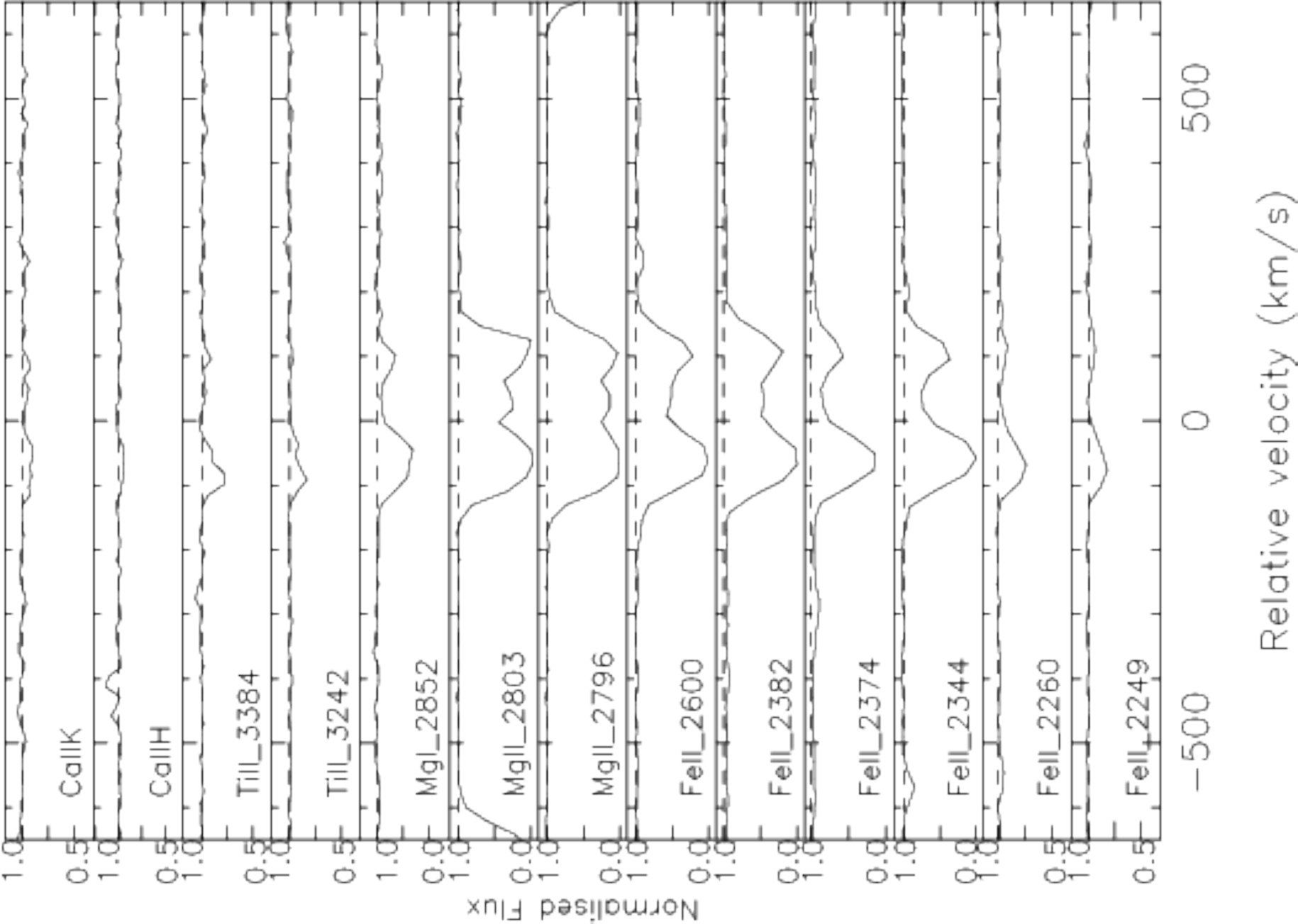}
\caption{{\bf Normalised X-Shooter 1D quasar spectrum of the absorption lines in the DLA  towards Q0452$-$1640.} \feii\ lines from $\lambda$ 2249 to 2600 \AA\ are seen as well as lines of other ions already reported by P\'eroux et al. (2008) at higher-resolution from VLT/UVES spectroscopy. In addition, we report both members of the \mgii\ doublet, 2 weak lines of \tiii\ ($\lambda \lambda$ 3242 and 3384) as well as the \caii\ H and K doublet which is detected in the NIR arm of the spectrum. 
}
\label{f:Q0452_Abs}
\end{center}
\end{figure}

\begin{figure*}
\begin{center}
\includegraphics[height=6.5cm, width=9.cm, angle=-90]{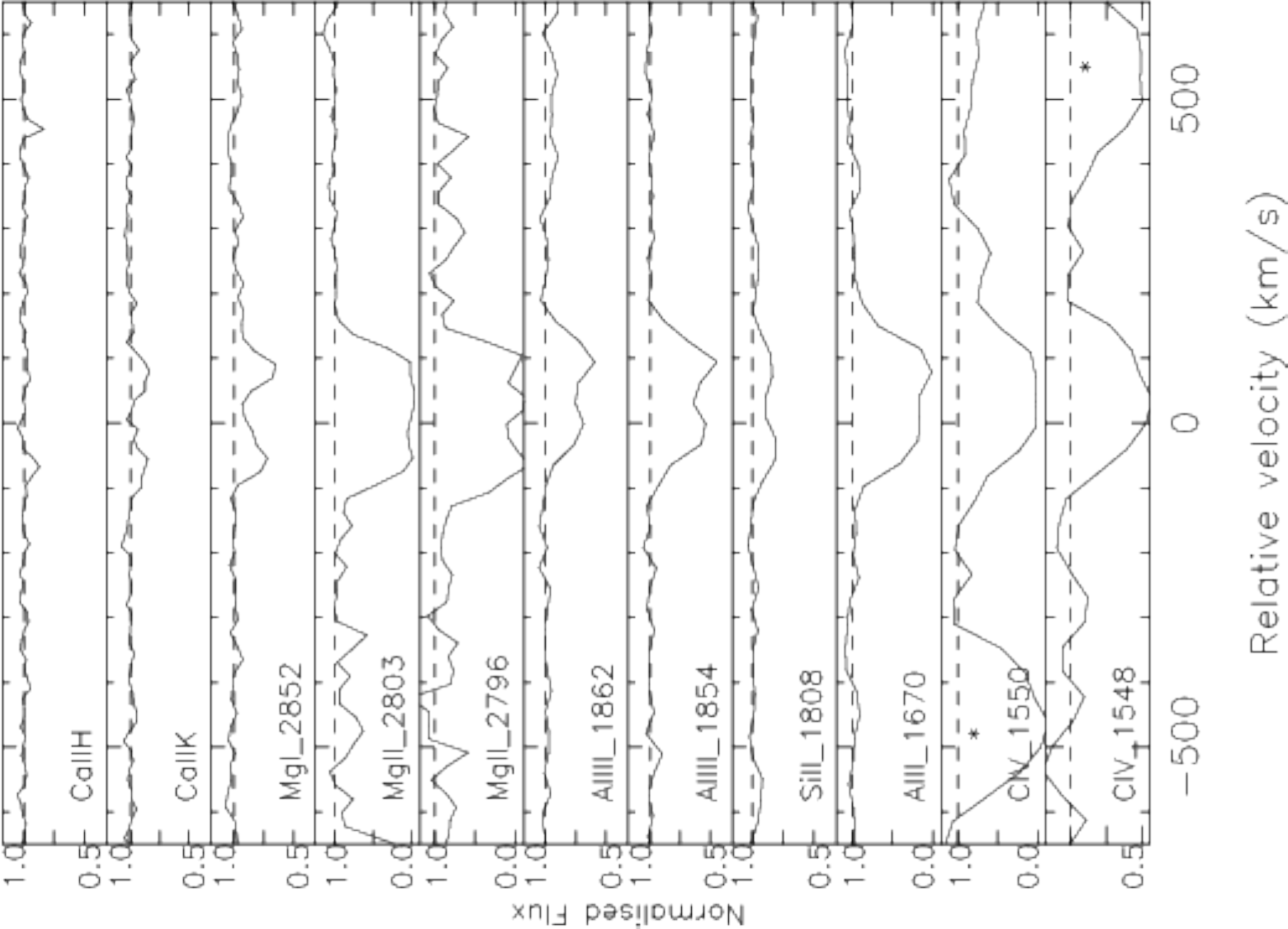}
\includegraphics[height=6.5cm, width=9.cm, angle=-90]{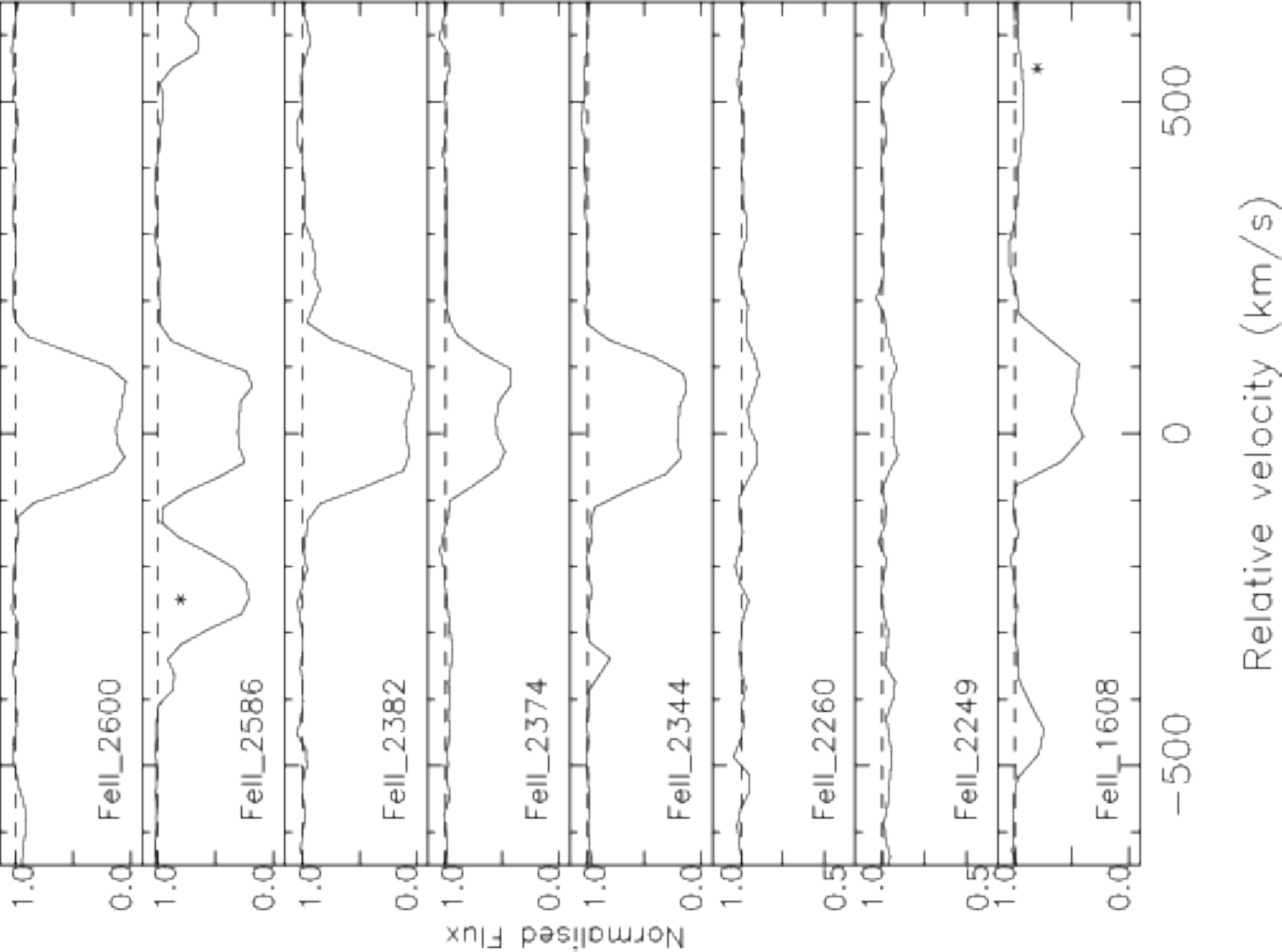}
\caption{{\bf Normalised X-Shooter 1D quasar spectrum of the absorption lines in the sub-DLA  towards Q2352$-$0028.} \feii\ lines from $\lambda$ 1608 to 2600 \AA\ are seen as well as other elements some of which are reported by Meiring et al. (2009) at higher-resolution from Magellan/MIKE spectroscopy. In addition, we report high-ionisation lines of the \civ\ doublet ($\lambda \lambda $1548, 1550) in the blue part of the X-Shooter spectrum, \siii\ ($\lambda $ 1808) as well as \alii\ ($\lambda$ 1670) and \aliii\ ($\lambda \lambda$ 1854, 1863). Finally, the \caii\ H and K lines are detected in the NIR arm of the spectrum. The features marked as '*' in panels \civ\ $\lambda \lambda$ 1548, 1550 and \feii\ $\lambda$ 2586 are features in the \lya\ forest lying near our target lines. }
\label{f:Q2352_Abs}
\end{center}
\end{figure*}

Voigt profile fits are commonly used to derive the column density of different elements detected in absorption in quasar spectra. We report here the results published in the literature of such analysis for the three targets under study as well as additional elements detected in the new X-Shooter spectra. The absorption line spectra presented in this study are used to identify additional lines (e.g. \aliii, \mgii, \caii\ H and K, etc) that were not identified in previous work. The abundances are given with respect to solar values using the convention [X/H]=$\log$(X/H)-$\log$(X/H)$_{\odot}$. Note that  molecular hydrogen H$_2$ in these absorbers is not covered by any of the X-Shooter spectra.

\underline {The DLA towards Q0302$-$223:} the absorber at \zabs=1.009 was first reported as a \mgii\ system by Petitjean \& Bergeron (1990), who also detected
\feii\ $\lambda$$\lambda$ 2586, 2600 and \mgi\ $\lambda$ 2852.  In this system, the \mgii\ absorption is a triple complex spanning
170 km/s. An HST Faint Object Spectrograph (FOS)
spectrum reveals several strong features from \cii, \civ, \ni,
\oi, \siii, \siiii\ and \siiv\ (Boiss\'e et al. 1998). Pettini \& Bowen (1997) used William Herschel Telescope (WHT) observations to study \znii\
and \crii\ lines, which were then followed by observations with the HIRES spectrograph at Keck (Pettini et al. 2000). These authors observe that two main groups of components at v=0 and v=--36 km/s, produce most of the absorption seen in this DLA.
Additional weaker components, at v=+35 and +121
km/s relative to \zabs=1.00945, are visible in the stronger
\feii\ lines. The resulting abundances with respect to solar are listed in Table~\ref{t:HI_Metals}. Finally, Boiss\'e et al. (1998) derive an upper limit on molecular hydrogen N(H$_2$), which translates into a molecular fraction f(H$_2$) $<$ 4.0 $\times$ 10$^{-3}$ (where f(H$_2$) = 2N(H$_2$)/[(2N(H$_2$) + N(HI)]).

In the new X-Shooter spectrum presented here, we detect a large number of ions, thanks to both the instrumental wavelength coverage (ranging from 1550 to 12300\AA\ in the absorber's rest-frame) and the high metallicity of the absorber. Figure~\ref{f:Q0302_Abs} presents a normalised spectrum of most of these lines. \feii\ lines from $\lambda$ 1608 to 2600 \AA\ are seen as well as the many Fe-peak elements already reported by Pettini et al. (2000) at higher-resolution from Keck/HIRES spectroscopy. We report 3 additional weak lines of \niii\ ($\lambda \lambda \lambda$ 1709, 1741, 1751) as well as \aliii\ ($\lambda$ 1670) and \aliii\ ($\lambda\lambda$ 1854 and 1862). Finally, the \caii\ H and K doublet is detected in the NIR arm of the spectrum.

\underline {The DLA towards Q0452$-$1640:}
Nestor et al. (2008) reported \znii\ and \crii\ detections in this \caii\ absorber using ISIS (Intermediate dispersion Spectrograph and Imaging System) on WHT (William Herschel Telescope). The quasar was first observed at high-resolution by P\'eroux et al. (2008) who then derived the metallicity of the absorber by fitting 13 components over $\sim$230 km/s to the absorption profile. The abundances for these systems are listed in Table~\ref{t:HI_Metals}. They also report possible detection of \coii\ in the same system, [Co/H]$=-$0.45. The \mgii\ doublet for that system falls in an UVES spectral coverage gap. 

In the new X-Shooter spectrum presented here, which covers 1550 to 12300 \AA\ in the absorber's rest-frame, we detect both the \mgii\ doublet and several lines of \tiii\ which were not reported. Figure~\ref{f:Q0452_Abs} presents a normalised spectrum of most of these lines. \feii\ lines from $\lambda$ 2249 to 2600 \AA\ are seen as well as other elements already reported by P\'eroux et al. (2008) at higher-resolution from VLT/UVES spectroscopy. In addition, we report the \caii\ H and K doublet which is detected in the NIR arm part of the spectrum.

\underline {The sub-DLA towards Q2352$-$0028:} Meiring et al. (2009) obtained a Magellan/MIKE spectrum of this quasar to study its \zabs=1.0318 sub-DLA. They use 11 components to fit the absorber over $\Delta$v $\sim$ 220 km/s. The corresponding abundances are listed in Table~\ref{t:HI_Metals}. The \mgii\ doublet is heavily saturated in this system and is used to derive a lower limit to the Mg abundance: [Mg/H]$>$$-$0.40. Note that this quasar sightline contains two other sub-DLAs: a system with \lognhi=19.18$\pm$0.08 at \zabs=0.8730 and another absorber with \lognhi=19.60$\pm$0.24 at \zabs=1.2467. 

In the new X-Shooter spectrum presented here, we detect a large number of new ions over the wavelength covered (1530-12200\AA\ in the absorber's rest-frame). Figure~\ref{f:Q2352_Abs} presents a normalised spectrum of most of these lines. \feii\ lines from $\lambda$ 1608 to 2600 \AA\ are seen as well as other elements some of which are reported by Meiring et al. (2009) at higher-resolution from Magellan/MIKE spectroscopy. In addition, we report high-ionisation lines of the \civ\ doublet ($\lambda \lambda$ 1548 and 1550) in the UVB arm of the X-Shooter spectrum, \siii\ ($\lambda $ 1808) as well as \alii\ ($\lambda$ 1670) and \aliii\ ($\lambda$ 1854 and 1862). Finally, the \caii\ H and K doublet is detected in the NIR arm of the spectrum.

%%%%%%%%%%%%%%%%%%%%%%%%
%%%%%%%%%%%%%%%%%%%%%%%%
\subsection{Integrated Ionised Phase Metallicity}

\subsubsection{Emission Line Measurements}

\begin{figure}
\begin{center}
\includegraphics[height=15.cm, width=8.5cm, angle=0]{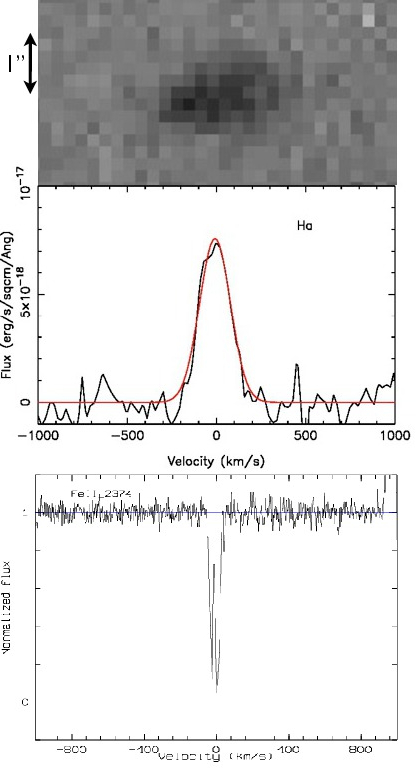}
\caption{{\bf Emission and absorption lines in the DLA towards Q0302$-$223.} In this and the following figures, the top panel is the 2D \ha\ emission line, the middle panel is the 1D \ha\ emission line and its gaussian fit both from the X-Shooter data and the bottom panel is the \feii\ $\lambda$ 2374 low-ionisation absorption profile from high-resolution spectroscopy (Keck/HIRES in this case, see Pettini et al. 2000). The velocity scale and zero point (matching \zabs) on the x-axis are the same for all panels. The quasar trace is not shown in these 2D images.}
\label{f:Q0302_2D_XSH}
\end{center}
\end{figure}

\begin{figure}
\begin{center}
\includegraphics[height=15.cm, width=8.5cm, angle=0]{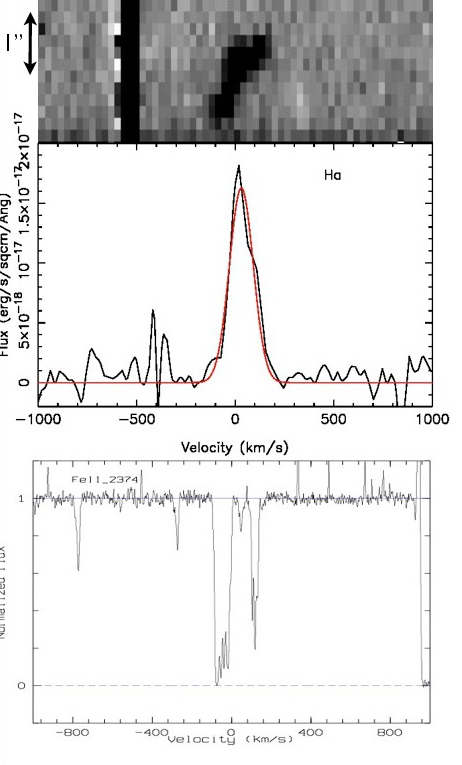}
\caption{{\bf Emission and absorption lines in the DLA towards Q0452$-$1640.} In this case, the rotation of the object is apparent from the 2D data. The black streak in the 2D image is a sky line. We also detect a clear double-peak profile in all the emission lines. The high-resolution spectrum is from VLT/UVES (P\'eroux et al. 2008).}
\label{f:Q0452_2D_XSH}
\end{center}
\end{figure}

\begin{figure}
\begin{center}
\includegraphics[height=15.cm, width=10.cm, angle=0]{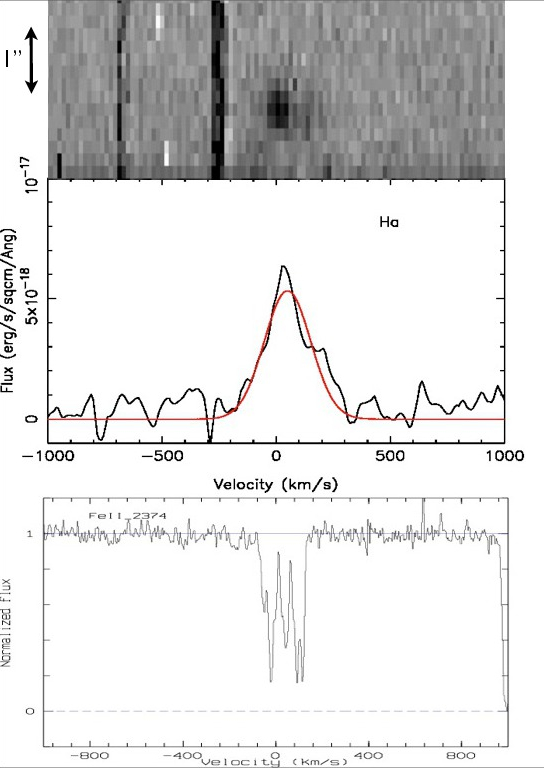}
\caption{{\bf Emission and absorption lines in the sub-DLA towards Q2352$-$0028.} The black streaks in the 2D image are sky lines. Fewer emission lines are detected in this case. The range of velocities spread by both the emission and the absorption profiles are comparable. The high-resolution spectrum is from Magellan/MIKE (Meiring et al. 2009).}
\label{f:Q2352_2D_XSH}
\end{center}
\end{figure}

In all three sets of data, the emission lines from the absorbing-galaxies are clearly detected in the VIS and NIR parts of the X-Shooter 2D spectra. This is illustrated for the predominant \ha\ lines in Figures~\ref{f:Q0302_2D_XSH}--\ref{f:Q2352_2D_XSH}, where the 2D spectra are plotted on the same velocity scale as the 1D spectra below. The zero point is set to the absorption redshift, \zabs, derived from the low-ionisation absorption profiles observed with high-resolution spectroscopy and shown in the bottom panels. We find that the absorption redshifts derived from the metallicity profiles are similar to the systemic redshifts measured from the emission lines which is remarkable given the impact parameters involved and rather unusual (Kacprzak et al. 2010a, 2011; Lundgren et al. 2012). The absorbing-galaxies are well-resolved and sufficiently offset from the quasar trace to be devoid of its contamination. Given the spectral resolution of X-Shooter in the VIS arm, the \oii\ doublet is resolved and the two separate components are visible on the 2D images. The continua of the absorbing-galaxies are not detected however.  We use gaussian fits to these emission lines to measure the integrated fluxes in erg/s/cm$^2$. Quoted errors reflect the uncertainties from the flux calibration, which is the dominant source of error in this case. The results are tabulated in Table~\ref{t:EmLines}. The last column of the table refers to the measure of \ha\ flux in the SINFONI data published by P\'eroux et al. (2011a) and P\'eroux et al. (2012). These fluxes are found to be consistent with the new X-Shooter measurements within the errors in all three cases.

\begin{figure*}
\begin{center}
\includegraphics[height=8.5cm, width=7.cm, angle=-90]{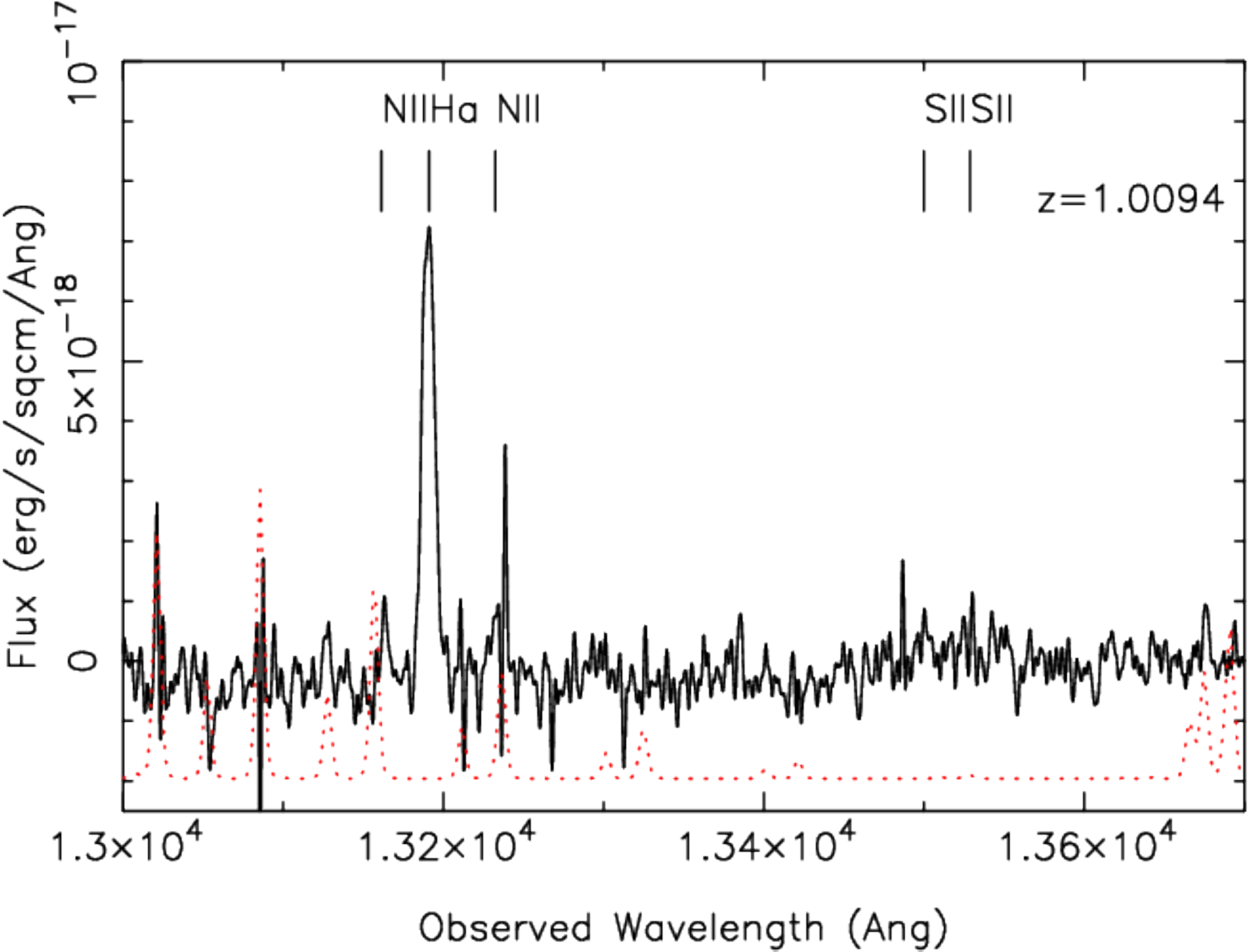}
\includegraphics[height=8.5cm, width=7.cm, angle=-90]{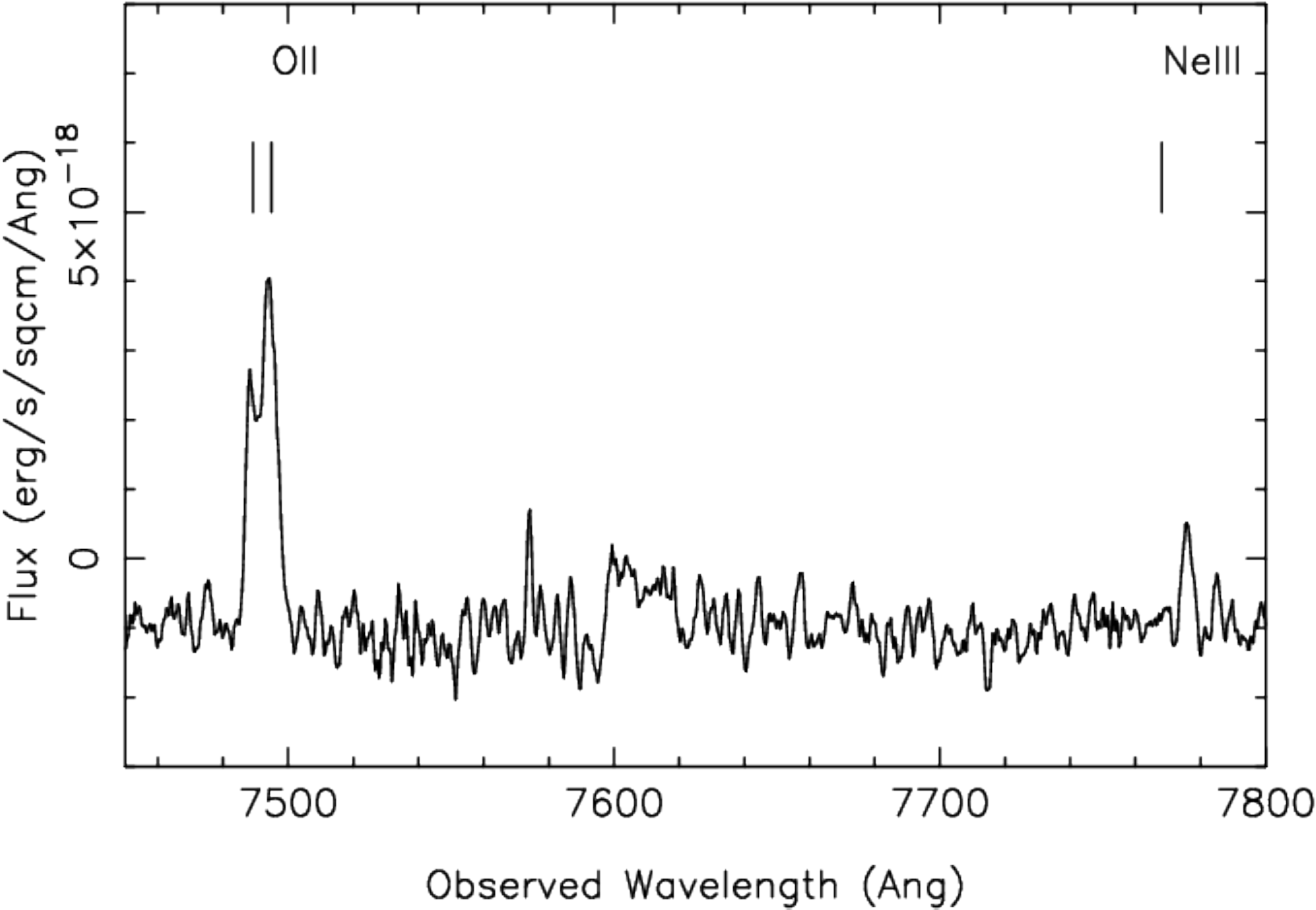}
\includegraphics[height=8.5cm, width=7.cm, angle=-90]{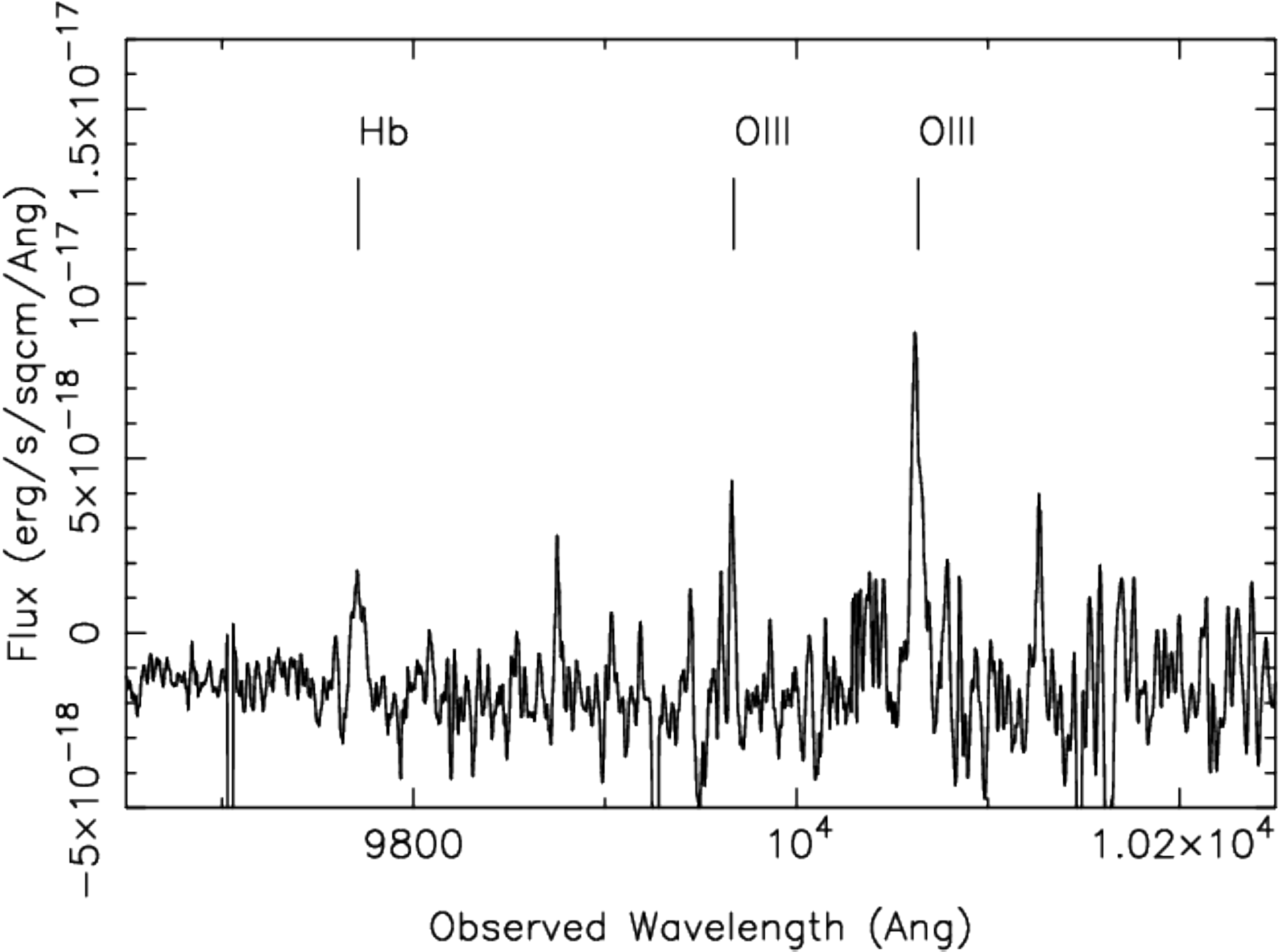}
\includegraphics[height=8.5cm, width=7.cm, angle=-90]{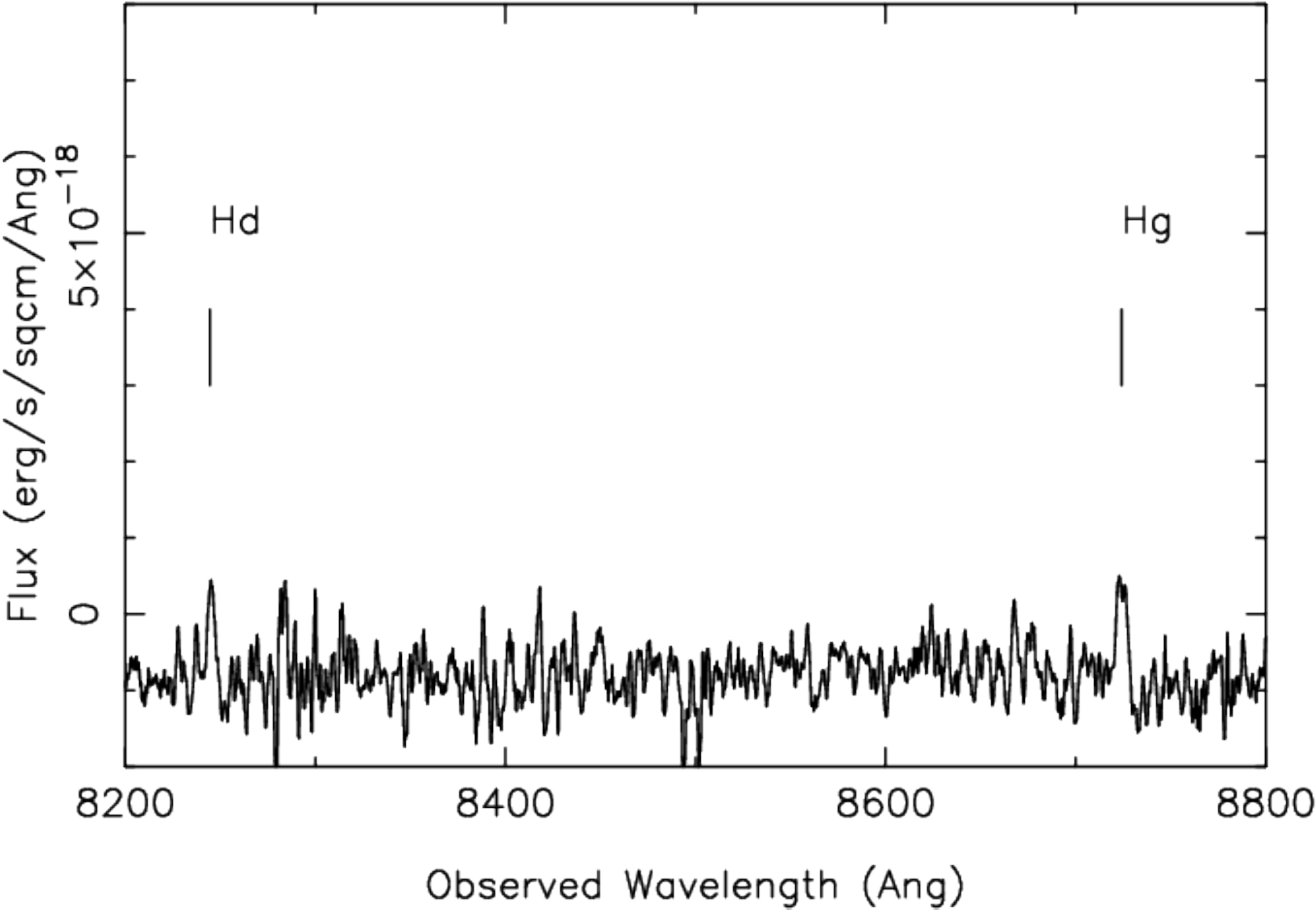}
\caption{{\bf Main emission lines detected in the X-Shooter spectrum of the DLA host towards Q0302$-$223.} In this figure and the following, in the panel for \ha, lying in the NIR arm, the dotted spectrum is the sky spectrum with arbitrary flux units, scaled for clarity, and indicating the position of the OH sky lines. Note that the \oii\ doublet is resolved at the X-Shooter resolution. The spectra are smoothed with a 5-pixel boxcar for display purpose only. In the case of the DLA towards Q0302$-$223, the \nii\ line of $\lambda$ 6550 is contaminated by a sky line and the \sii\ doublet and \nii\ $\lambda$ 6585 are not detected.
 }
\label{f:Q0302_1D_XSH}
\end{center}
\end{figure*}

\underline {The DLA towards Q0302$-$223:} many different emission lines are detected in the X-Shooter absorbing-galaxy spectrum of this DLA, including \ha, both lines of \oiii\ (4960.29 and 5008.24 \AA), \hb, \hg, \neiii\ (3870.91), \hd\ and the \oii\ doublet. Figure~\ref{f:Q0302_1D_XSH} shows portions of the spectra with some of these lines. We note that the detection of \hg\ emission at such redshift is rare. The \nii\ line of $\lambda$ 6550 is contaminated by a sky line. The results of the emission lines fits are tabulated in Table~\ref{t:EmLines} which also includes 3-$\sigma$ limits for non-detections of the \sii\ doublet and \nii\ $\lambda$ 6585 in this case.

\begin{figure*}
\begin{center}
\includegraphics[height=8.5cm, width=7.cm, angle=-90]{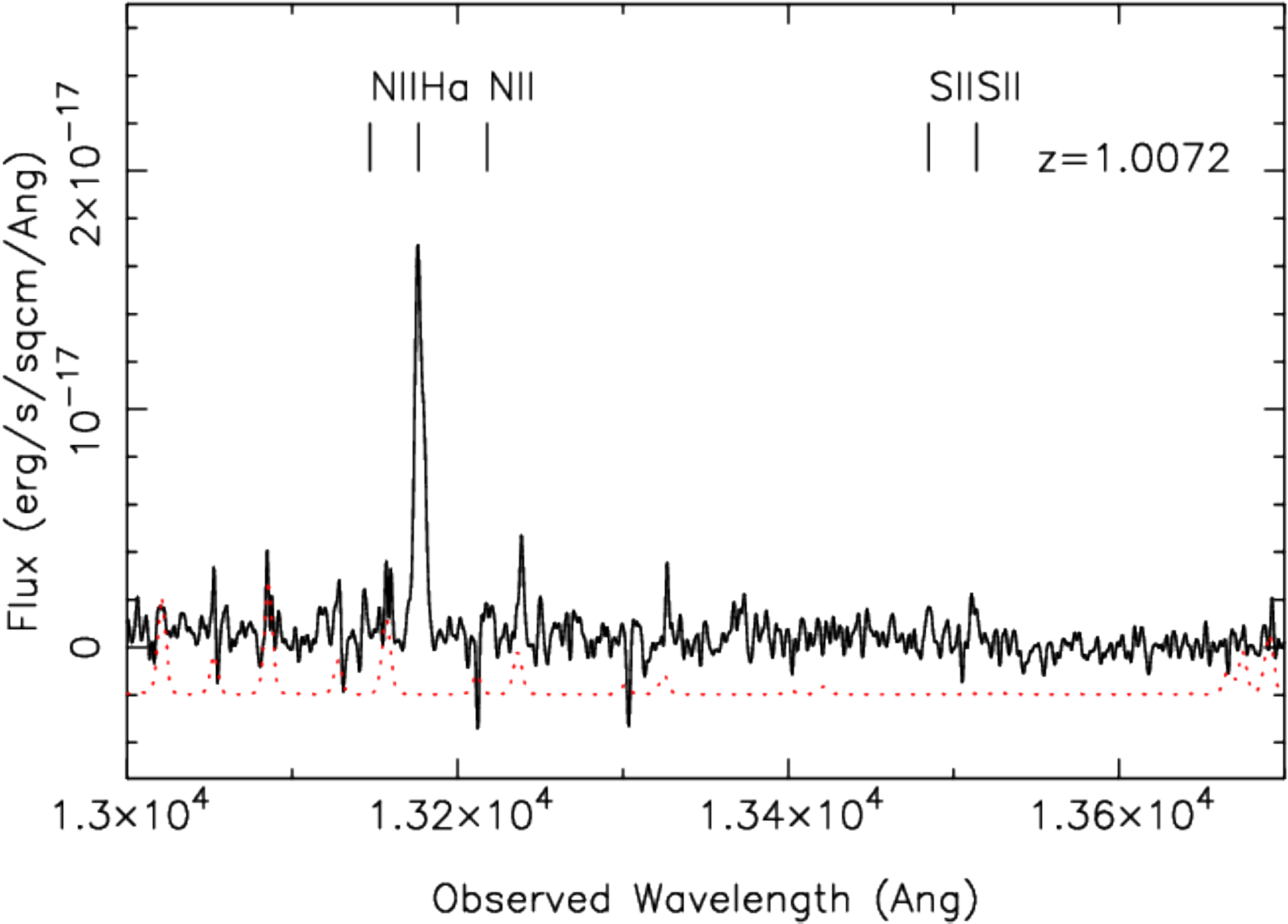}
\includegraphics[height=8.5cm, width=7.cm, angle=-90]{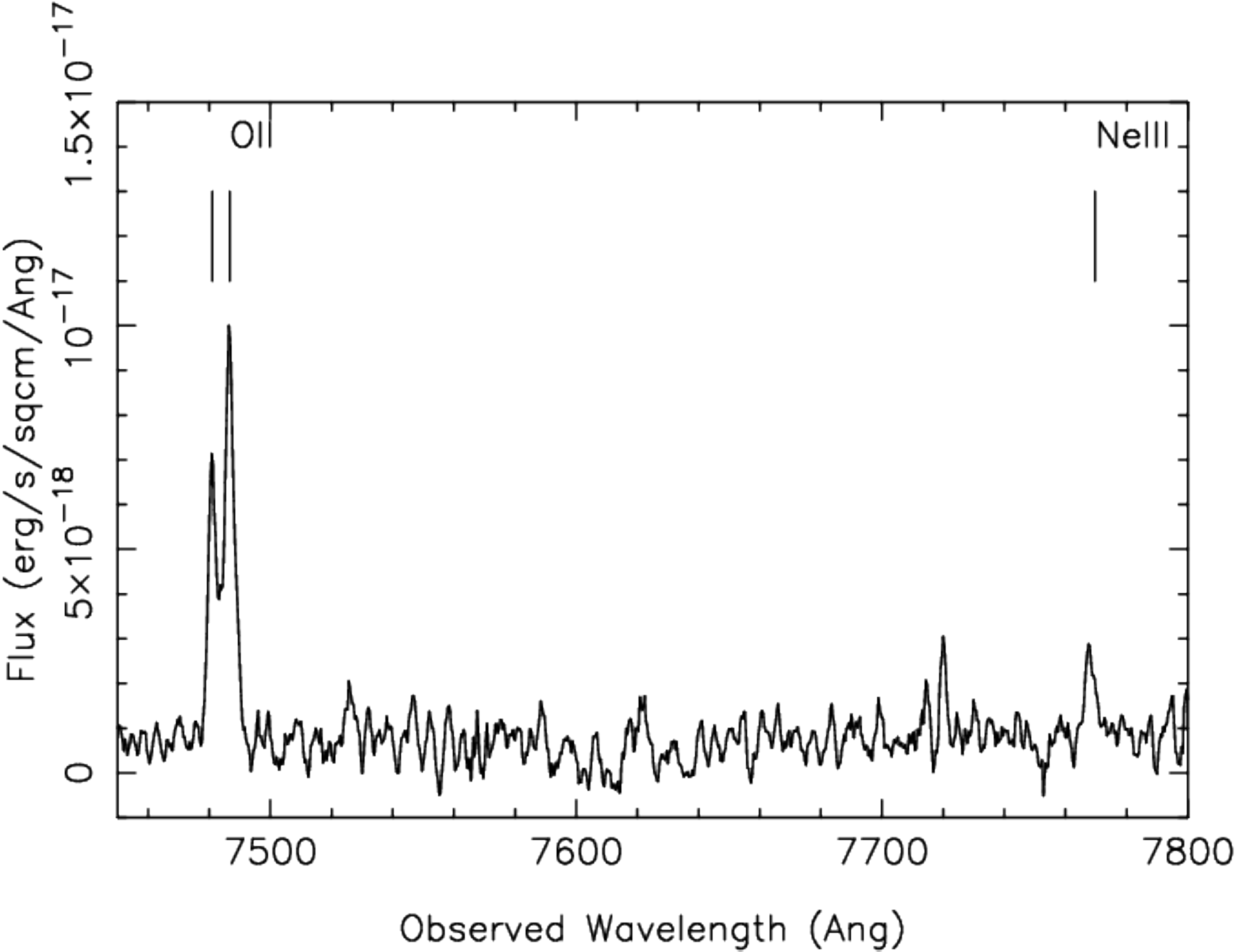}
\includegraphics[height=8.5cm, width=7.cm, angle=-90]{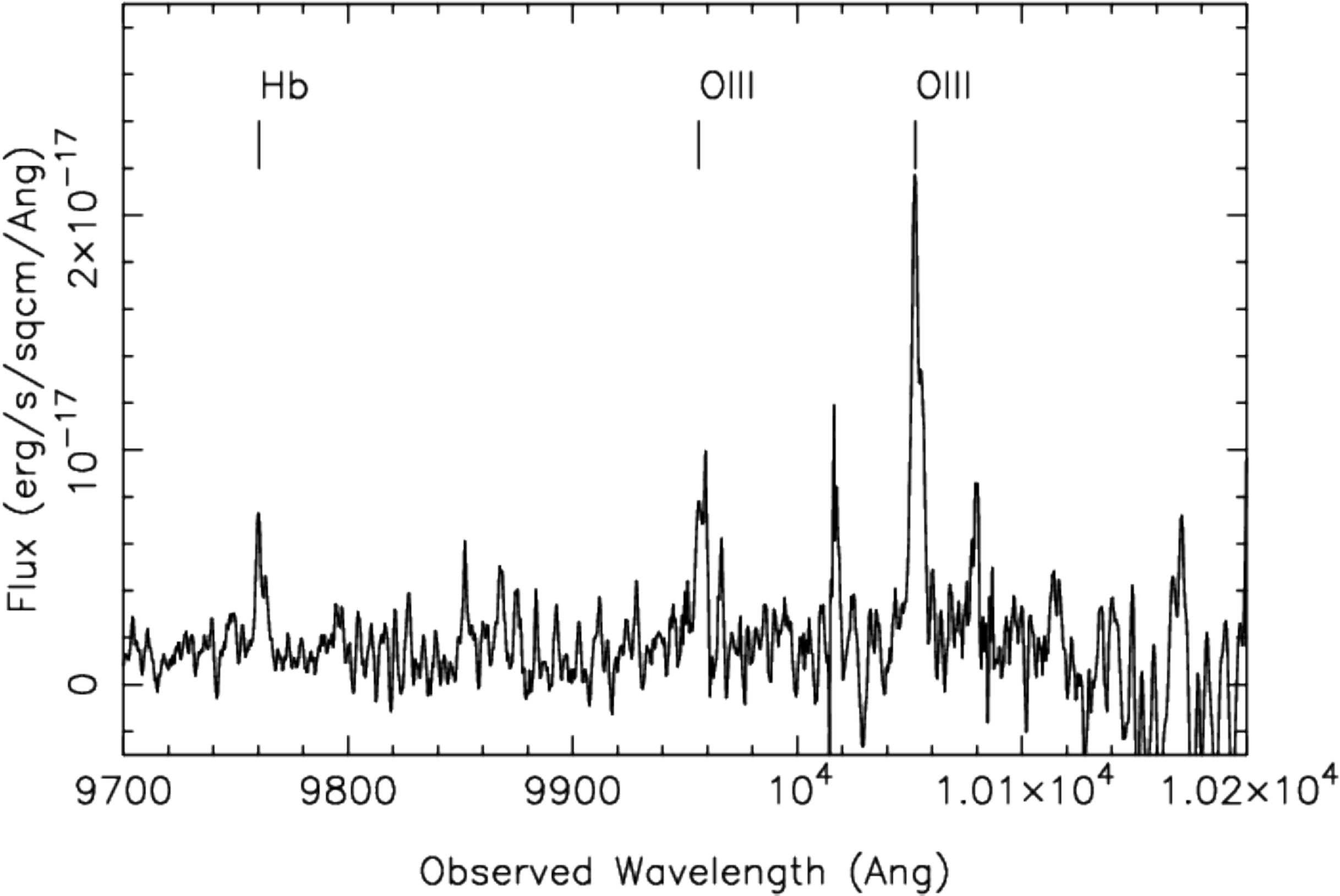}
\includegraphics[height=8.5cm, width=7.cm, angle=-90]{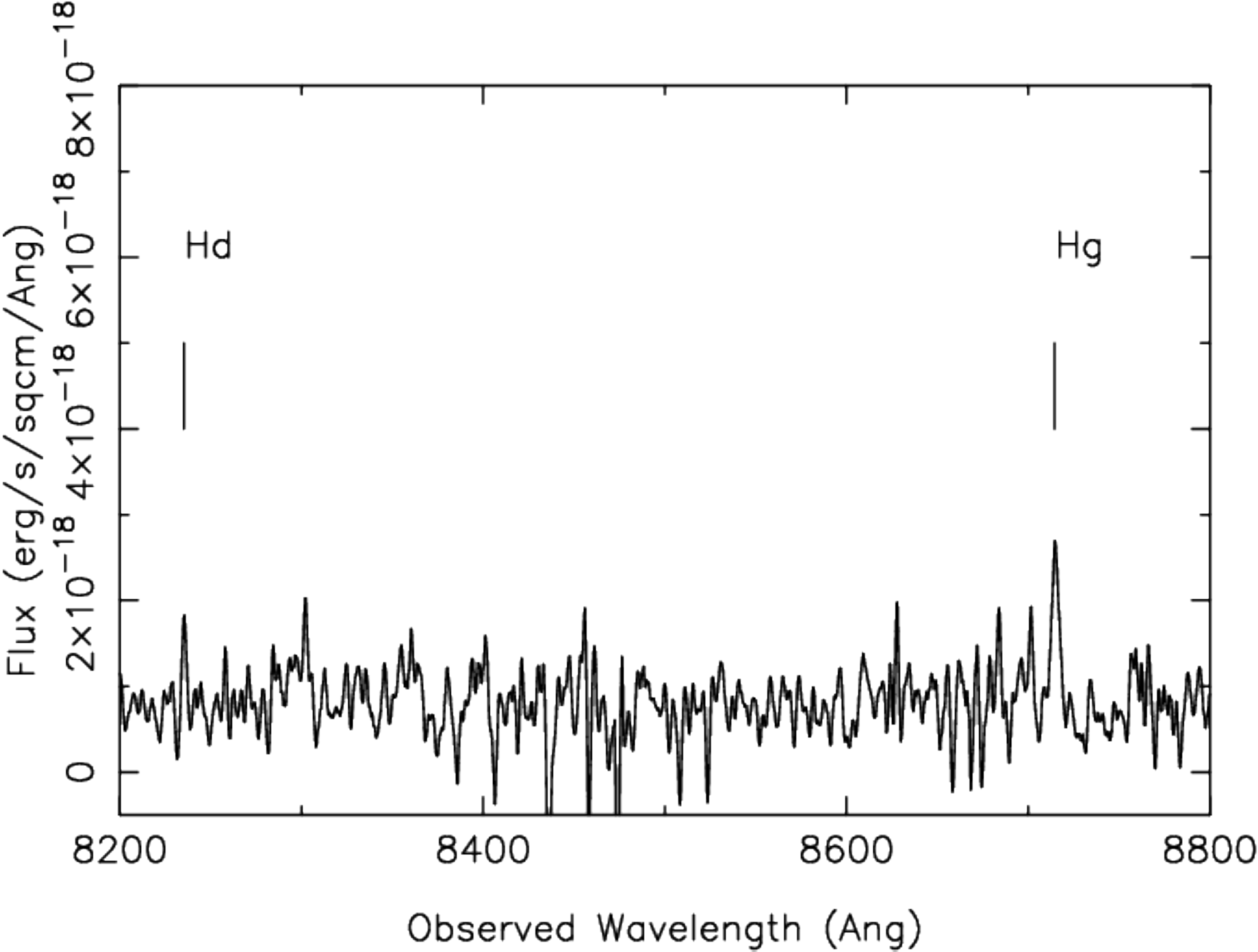}
\caption{{\bf Main emission lines detected in the X-Shooter spectrum of the DLA host towards Q0452$-$1640.} All the emission lines covered by the X-Shooter spectrum are detected with the exception of \nii\ $\lambda$ 6550. Given the rms of the spectra, the detection of \nii\ $\lambda$ 6585 is consistent with the non-detection of \nii\ $\lambda$ 6550 assuming a ratio \nii\ $\lambda$ 6585/\nii\ $\lambda$ 6550=3 (Storey \& Zeippen 2000). The detections therefore include \sii, \ha, \nii\, both lines of \oiii, \hb, \hg, \neiii, \hd\ and the \oii\ doublet.}
\label{f:Q0452_1D_XSH}
\end{center}
\end{figure*}

\underline {The DLA towards Q0452$-$1640: } in this system, all the strongest emission lines covered by the X-Shooter spectrum are detected with the exception of \nii\ $\lambda$ 6550. The detections therefore include \sii, \ha, \nii\ $\lambda$ 6585, both lines of \oiii, \hb, \hg, \neiii, \hd\ and the \oii\ doublet. Again, we note the rare detection of \hg\ emission at such high redshift.  Given the rms of the spectra, the detection of \nii\ $\lambda$ 6585 is consistent with the non-detection of \nii\ $\lambda$ 6550 assuming a ratio \nii\ $\lambda$ 6585/ \nii\ $\lambda$ 6550=3 (Storey \& Zeippen 2000). Figure~\ref{f:Q0452_1D_XSH} shows portions of the spectra with some of these emission lines. The results of the fits are tabulated in Table~\ref{t:EmLines}.

\underline {The sub-DLA towards Q2352$-$0028:} few lines are detected in the X-Shooter spectrum of this absorbing-galaxy. The only reliable detections are from \ha\ and the \oii\ doublet. Interestingly, the line of \nii\ $\lambda$ 6585 is significantly detected in the SINFONI data but is not seen in the X-Shooter spectrum. Given the rms of the spectra, the non-detection of \oiii\ $\lambda$ 4960 is consistent with the non-detection of \oiii\ $\lambda$ 5008 assuming a ratio \oiii\ $\lambda$ 5008/ \oiii\ $\lambda$ 4960=3 (Storey \& Zeippen 2000). The other non-detections lead to upper limits which are tabulated in Table~\ref{t:EmLines}.

\subsubsection{\hii\ Abundances}

\begin{table*}
\begin{center}
\caption{{\bf Absorbing-galaxy emission line properties.} Emission lines covered by the X-Shooter spectra are listed for each of the three systems. The rest vacuum wavelengths are listed for each line. The fluxes are expressed in units of $\times$ 10$^{-17}$ erg/s/cm$^2$. The last column of the Table refers to the measure of \ha\ flux in the SINFONI data published by P\'eroux et al. (2011a) and P\'eroux et al. (2012). These fluxes are found to be consistent with the new X-Shooter measurements within the errors in all three cases. 
}
\label{t:EmLines}
\begin{tabular}{cccccccc} 
\hline\hline             
Quasar  &\oii\	&\oii\	&\neiii\ &\hd\ &\hg\ &\hb\ &\oiii\ \\
&3727.09&3729.87&3870.91&4102.89&4341.68&4862.68&4960.29\\
\hline
Q0302$-$223  &$2.0\pm1.0$ &$3.0\pm1.5$ &$0.5\pm0.2$ &$0.7\pm0.3$ &$0.9\pm0.4$ &$2.3\pm1.1.$ &$2.0\pm1.0$ \\
Q0452$-$1640 &$1.9\pm0.6$ &$3.8\pm1.1$ &$0.9\pm0.3$ &$0.4\pm0.1$ &$0.9\pm0.3$ &$1.7\pm0.5$ &$2.4\pm0.7$ \\
Q2352$-$0028 &$0.8\pm0.4$ &$0.2\pm0.1$ &$<1.5$ &$<1.5$ &$<1.5$ &$<3.0$ &$<3.0$ \\
\hline\hline

Quasar&\oiii\ &\nii\ &\ha\ &\nii\ & \sii\ &\sii\ &SINFONI \ha\\\
&5008.24&6549.86&6564.61&6585.27&6718.29&6732.67 &6564.61\\
\hline
Q0302$-$223&$6.8\pm3.4$ &$<1.2$ &$7.4\pm3.7$ &$<1.2$ &$<1.2$ &$<1.2$&7.7$\pm$2.7 \\
Q0452$-$1640&$7.3\pm2.2$ &$<3.0$ &$10.8\pm3.2$ &$1.6\pm0.5$ &$0.7\pm0.2$ &$0.4\pm0.1$ &14.5$\pm$4.3\\
Q2352$-$0028&$<3.0$ &$<1.5$ &$5.5\pm2.7$ &$<1.5$ &$<1.5$ &$<1.5$&4.9$\pm$2.4 \\
\hline\hline
\label{t:EmLines}
\end{tabular}
\end{center}			       			 	 
\end{table*}

Emission lines are commonly used to estimate gas-phase metallicities in extragalactic \hii\ regions. One of the most accurate estimates requires that the electron temperature be measured for each object. The temperature can be calculated using the ratio of the auroral \oiii\ $\lambda$ 4363 to the nebular \oiii\ $\lambda\lambda$ 4960, 5008 lines (Osterbrock 1989). Auroral lines are intrinsically faint and have only recently been detected in high-redshift galaxies at low confidence levels (Yuan \& Kewley 2009; Christensen et al. 2012). Other metallicity measurements are available whereby one does not need the electron temperature, including two indices which we can measure with our data: N2 and O3N2. The N2 index uses the \nii\ $\lambda$ 6583/\ha\ ratio to estimate the oxygen abundance (see Storchi-Bergmann et al. 1994). The N2 index saturates at solar metallicity so that Pettini \& Pagel (2004) have proposed the O3N2 index which combines N2 with measurements of \oiii\ $\lambda$ 5008 and \hb, providing a more accurate index for super-solar metallicities. These ratios have been calibrated as oxygen abundance indicators in local \hii\ regions. The last indicator used is the R$_{23}$ indicator which was first introduced by Pagel et al. (1979), and is widely used for measuring local metal abundances if the fluxes of \oiii\ and \oii\ are known:  

\begin{itemize} 

\item N2=$\log$[F(\nii--6585)/F(\ha)]

\item O3N2=\\
$\log$[(F(\oiii--4960)+F(\oiii--5008))/F(\hb)]/[F(\nii--6585)/F(\ha)]

\item R$_{23}$=\\
$\log$ (F(\oii--3727) + F(\oiii--3729) + F(\oiii--4960) + F(\oiii--5008) / F(\hb) 

\end{itemize}

The first two indicators have been calibrated using direct O/H measurements for local \hii\ regions by Pettini \& Pagel (2004). These authors have established a linear relation to the metallicity of the form (see also P\'erez-Montero \& Contini 2009):

\begin{itemize} 

\item 12+$\log$ (O/H)=8.90+0.57 $\times$ N2

\item 12+$\log$ (O/H)=8.73$-$0.32 $\times$ O3N2

\end{itemize}

These relations are valid for 7.50$<$12+$\log$ (O/H)$<$8.75 and 8.12$<$12+$\log$ (O/H)$<$9.05, respectively. We have not corrected the N2  abundance for dust extinction as this indicator uses lines which are close in wavelength, so that the effects of the dust extinction on this ratio is small. For O3N2 and R$_{23}$, we use our estimates of $E(B-V)$ for these absorbing-galaxies (see section 4.2) to calculate the Milky Way-type extinction-corrected abundances.

Furthermore, we used the relation between R$_{23}$ and metallicity based on the work of Kobulnicky \& Kewley (2004). These authors used a set of ionisation parameters and oxygen abundance diagnostics. The local ionisation state in an \hii\ region is often characterised by the ionisation parameter, $q$, which is the ratio of the ionising photon flux and the number density of hydrogen atoms. Commonly, ionisation parameters are estimated using O23 = $\log$[(\oiii\ $\lambda $4960+\oiii\ $\lambda $5008)/(\oii\ $\lambda$ 3727 + \oii\ $\lambda$ 3729)] (equation 13 in Kobulnicky \& Kewley 2004). Here, we derive O32=+0.2$\pm$0.2, -0.2$\pm$0.1 and $<$+1.2 corrected for extinction for the absorbing-galaxies towards Q0302$-$223, Q0452$-$1640 and Q2352$-$0028, respectively. Lilly, Carollo \& Stockton (2003) showed that a vast majority of local objects have a value of O32 $<$ 1, while the available data for objects at z $>$ 2 indicate O32 $>$ 1. Indeed, to date, only a small sample of O32 measurements have been obtained for objects at z$>$2 (Pettini et al. 2001; Lemoine-Busserolle et al.
2003; Maiolino et al. 2008). Our values are more typical of the star-forming galaxies at intermediate redshift (0.47 $<$ z $<$ 0.92)
from Lilly, Carollo \& Stockton (2003). Our results at z$\sim$1 are therefore in line with the low-redshift values reported in the literature.

Kobulnicky \& Kewley (2004) also provide an analytical fit to relate the R$_{23}$-metallicity relation for the high-metallicity upper branch (their equation 15) and for the lower branch (their equation 16). We derive R$_{23}$=0.8$\pm$0.2 and 1.1$\pm$0.2 for Q0302$-$223 and Q0452$-$1640, respectively. We iteratively solve these equations to estimate $q$ and metallicity for the upper and lower branches. For both objects, we find the lower branch solutions to be inconsistent with other metallicity indicators and therefore favour the upper-branch metallicity estimates. The resulting \hii\ metallicity and ionisation parameter, $q$, for the three absorbing-galaxies presented in this paper (Q0302$-$223, Q0452$-$1640 and Q2352$-$0028) are tabulated in Table~\ref{t:HII_Metals}. The table lists a reference to the value from N2-index as derived from previous SINFONI observations (P\'eroux et al. 2011b, 2012), the results from three different metallicity indicators described here (some of which are limits), the mean of these values, the metallicity in the neutral phase measured in absorption with the nearly undepleted element Zn along the line-of-sight to the background quasar, the ionisation parameter, $q$ and the dust extinction $E(B-V)$ deduced from the Balmer decrement, \ha/\hb\ (see following sections). For Q0452$-$1640, our results show that the three different metallicity indicators lead to consistent results. We therefore compute the mean of the various metallicity indicators obtained from X-Shooter observations. In the case of Q0302$-$223, we only kept the results from the R$_{23}$ index given that it is the only measure (as opposed to limits) and it is consistent with other indicators within the error estimates. These results are tabulated in Table~\ref{t:HII_Metals}. We find that the new N2 values derived from X-Shooter and the previous SINFONI N2-index-based measurements agree within the errors.

\begin{table*}
\begin{center}
\caption{{\bf Metallicities with respect to solar in units of 12+log(O/H)}. The table lists a reference to the value from N2-index as derived from previous SINFONI observations (P\'eroux et al. 2011b, 2012), the results from three different metallicity indicators (some of which are limits), the mean of these values, metallicity in the neutral phase measured in absorption with the undepleted Zn element along the line-of-sight to the background quasar, the ionisation parameter, $q$, derived from the the extinction-corrected ratio \oiii\ 5008/ \oii\ 3727 and the dust extinction $E(B-V)$ deduced from the Balmer decrement, \ha/\hb.}
\label{t:HII_Metals}
\begin{tabular}{ccccccccccc}
\hline\hline
Quasar 		&\hii\ Metallicity  &N2  &O3N2 &R$_{23}$ &\hii\ Metallicity  &\hi\ Metallicity &Ioni. Parameter &$E(B-V)$\\
&N2 (SINFONI)&(X-Shooter)&&&(X-Shooter mean)&(absorption) &$q$ [cm/s] &\\
\hline
Q0302$-$223   	&$<$8.6 			&$<$8.45		&$>$8.29		&8.73$\pm$0.23	&8.7$\pm$0.2 	&8.15$\pm$0.12 	       &0.7 $\times$ 10$^8$		&0.1$\pm$0.8\\	
Q0452$-$1640 	&8.4$\pm$0.1        &8.43$\pm$0.06	&8.22$\pm$0.04	&7.96$\pm$0.04	&8.2$\pm$0.2		&7.70$\pm$0.08 	       &0.2 $\times$ 10$^8$	         &0.8$\pm$0.4\\	
Q2352$-$0028 	&8.4$\pm$0.3      	&$<$8.58		&--				&--    			&$<$8.6	    
 &$<$8.15 &$<$4 $\times$ 10$^8$   &$>$$-$0.4\\	
\hline\hline 				       			 	 
\end{tabular}			       			 	 
\end{center}			       			 	 
\end{table*}

\subsubsection{AGN Contamination}

The nebular emission line ratios can also probe the ionisation by an AGN, by comparing the line ratios with, for example, a Baldwin-Philips-Terlevich (hereafter BPT) diagram (Baldwin, Phillips \&Terlevich 1981). This diagram separates
star-forming galaxies and AGN according to the optical line
ratios \nii\ $\lambda$ 6583/\ha\ and \oiii\ $\lambda$ 5008/\hb. The star-forming galaxies fall in a
region of lower \nii\ $\lambda$ 6583/\ha\ for a range of values of \oiii\ $\lambda$ 5008/\hb. The bordering line between the star-forming and AGN regions are defined from  Kauffmann et al. (2003)
that demarcates star-forming galaxies and AGN on an empirical
basis, and from Kewley et al. (2001a) which represents a
limit to the line-flux ratios that can be produced for star-forming
regions from photoionisation plus stellar population synthesis
models.
Recently, Shapley et al. (2005) and Erb et al. (2006a) have presented evidence indicating a difference
between \hii\ regions in high-redshift galaxies and those in local
galaxies. They demonstrate that a fraction of the high-redshift star-forming galaxies at z$\sim$1--2 lies offset from the local population
of \hii\ regions and star-forming galaxies, displaced toward
higher \nii\ $\lambda$ 6583/\ha\ and \oiii\ $\lambda$ 5008/\hb\ values. 

Out of the three objects presented here, one has a \nii\ $\lambda$ 6583 detection and the other two lead to lower limits. We derive $\log$(\nii\ $\lambda$ 6583/\ha) $<-$0.8, =$-$0.8$\pm$0.1 and $<-$0.6 for the absorbing galaxies towards Q0302$-$223, Q0452$-$1640 and Q2352$-$0028, respectively. 
Similarly, both the \oiii\ $\lambda$ 5008 and \hb\ lines are not detected in absorbing-galaxy towards Q2352$-$0028. We derive $\log$[(\oiii\ $\lambda$ 4960 + \oiii\ $\lambda$ 5008)/\hb] = 0.6$\pm$0.2 and 0.7$\pm$0.1 for the absorbing galaxies towards Q0302$-$223 and Q0452$-$1640, respectively. We find that all three objects do not show contribution from AGNs and prove to be consistent with starbursts. Even for Q2352$-$0028 which only has a limit from the \nii\ $\lambda$ 6583/\ha\ ratio, the result indicates that it is consistent with expectations for
star-forming galaxies with no apparent AGN contamination or shock ionisation.

%%%%%%%%%%%%%%%%%%%%%%%%
%%%%%%%%%%%%%%%%%%%%%%%%
\subsection{Analysis}

\subsubsection{Internal Gradients}

\noindent

In the local Universe, disc galaxies almost invariably
show negative metallicity gradients, with the central abundances being higher than the one in the outskirts of the galaxy. The metallicity gradients in local galaxies are strongly correlated with Hubble type, bar strength, and merging events, with
more flattened gradients for barred galaxies and merging pairs
(Kewley et al. 2010; Rupke, Kewley \& Chien 2010). 

The small sizes of high-redshift galaxies
(Law et al. 2007; Conselice et al.
2008; Smail et al. 2007; Swinbank et al. 2010) mean that
sub-kpc resolution observations would be preferred to measure the
abundances reliably. To overcome this problem, spatially resolved measurements
of \nii/\ha\ have been made in gravitationally amplified
galaxies at z=1.5--2.5. Jones et al. (2010) measure strong, negative
gradients at z$\sim$2 suggesting that if z $\sim$2 starbursts evolve
into spiral discs today, then the gradients must flatten by 0.05--0.1 dex/kpc over the last 8--10Gyr. Furthermore, the high-redshift spiral galaxy metallicity gradient of Yuan et al. (2011) is consistent with 'inside-out' disc formation model. More recently, Swinbank et al. (2012) have reported metallicity
gradients in unlensed systems at 0.84$<$z$<$2.23 which are negative or consistent with zero, with
an average $-$0.027$\pm$0.005 dex/kpc,
which is comparable with the slope seen in the thick disc of the
Milky-Way. 
The authors find that all three isolated galaxies have negative gradients, while the merging
system has a positive gradient. 

In fact, only a few positive gradients have been measured, all of which are relatively shallow ($<+$ 0.05 dex/kpc). Positive
metallicity gradients have been claimed for some high
redshift galaxies, most of them with no signs of recent
interaction or merging. Cresci et al. (2010) and Queyrel et al. (2012) have suggested they could be caused by high accretion rates of
metal-poor gas onto the nucleus which occur only at high-redshift or during mergers. Queyrel et al. (2012) argue that these  inflows are preferably due to interactions whereas Cresci et al. (2010) relate them to cold flows. Queyrel et al. (2012) also claim evidence for gas inflow and metallicity gradient flattening
in close pairs and interacting galaxies. These authors point to a transition period at z$\sim$1--2 between 
cold accretion (at high-redshift) and mergers (at lower-redshift)  in the growing process of massive galaxies. These reported 'inverted' gradients are puzzling since we know it is difficult to form a positive
gradient given the fraction of gas converted into stars
per unit time must increase with radius, while the dynamical time-scale generally decreases with radius. The
corollary is that a higher fraction of metal-enriched gas
must be lost to outflows at small radii. Positive gradients
for interacting systems may be a signature of the redistribution
of the metal-rich gas produced from a central starburst
(Werk et al. 2010; Rupke, Kewley \& Chien 2010), or possibly as a result
of inflow of relatively unenriched gas from the halo (or IGM)
if it is able to intersect the central regions of the disc galaxy
without being disrupted. Another explanation
is an increasing contribution of shocks at large radius
rather than a positive metallicity gradient (Kewley et al. 2010).

We now know how merging processes affect the metallicity gradients: the gradients flatten immediately after first close pass (Barton, Geller \& Kenyon 2003; Kewley, Geller \& Barton 2006). The metallicity in mergers can be used as a signpost of first pericentre passage (Kewley et al. 2010). Models of Kobayashi (2004) predicts that in the case of monolithic collapse, there will be steep metallicity gradient which shows no time evolution. On the contrary, major mergers are expected to flatten the metallicity gradient with signs of an evolution with time. Therefore, the gradient evolution with redshift is related to the radial size growth predicted by models. Recently, Swinbank et al. (2012) reproduced their observed metallicity gradients with predictions for the gas discs of star-forming
galaxies in the cosmologically based hydrodynamic
simulations. In these models, the inner disc undergoes initial
rapid collapse and star-formation with gas accretion (from the halo and/or IGM) depositing relatively unenriched
material at the outer disc, causing negative abundance gradients.
At lower redshift (when gas accretion from the IGM
is less efficient) the abundance gradients flatten as the outer
disc becomes enriched by star-formation (and/or the redistribution
of gas from the inner disc). We note however that M101 is an important exception to this description, being at low-redshift but with a strong gradient (Kennicutt, Bresolin \& Garnett 2003).

The systems discussed in this paper are part of a larger group of galaxies observed with SINFONI. For three of the absorbing-galaxies observed in that sample, we have built 2D metallicity maps based on \nii\ detection. The map of the system towards Q1009$-$0026 is presented in P\'eroux et al. (2011a) and the ones for Q0452$-$1640 and Q2352$-$0028 studied here are presented in P\'eroux et al. (2013a). We derive internal metallicity gradients of $-$0.11$\pm$0.17, $<$$-$0.10 and $-$0.07$\pm$0.35 dex/kpc for Q0452$-$1640, Q1009$-$0026 and Q2352$-$0028, respectively. Therefore, the internal gradients of metallicity with radius of these three objects are flat or at most weakly negative for two of the three systems. 

These estimates are in line with the gradients reported in the local Universe. For the local
galaxy gradients, the isolated spiral control sample of
Rupke, Kewley \& Chien (2010) indicates a median slope of $-$0.041$\pm$0.009 dex/kpc. The median slope of gradients from that sample is consistent with
the values of Zaritsky, Kennicutt \& Huchra (1994) and van Zee et al. (1998)
samples. The Zaritsky, Kennicutt \& Huchra (1994) gradient sample is composed
of 39 local disc galaxies, and the van Zee et al. (1998) sample
added another 11 local spirals. Local early-type galaxies have
even shallower gradients (Henry \& Worthey 1999). Therefore, our observations resemble more the local samples or the results of Queyrel et al. (2009) at z$\sim$1.2 than the gradient recently reported in lensed galaxies at z$\sim$2 (Jones et al. 2010; Yuan et al. 2011; Jones et al. 2012). 

Prantzos \& Boissier (2000) modeled metallicity gradients of
spiral galaxies for different
velocities (V$_c$) and spin parameters ($\lambda$). We find that the metallicity gradients reported here are comparable with the predictions by Prantzos \& Boissier (2000)
for V$_c$ = 220 km/s which is typical of the dynamics of these systems (see P\'eroux et al. 2013a). At z $\sim$1,
the model disc galaxies from the GIMIC simulation presented by Sinwbank et al. (2012) display
negative metallicity gradients which are also consistent with our
observations. 

\begin{figure}
\begin{center}
\includegraphics[height=8.5cm, width=7.cm, angle=-90]{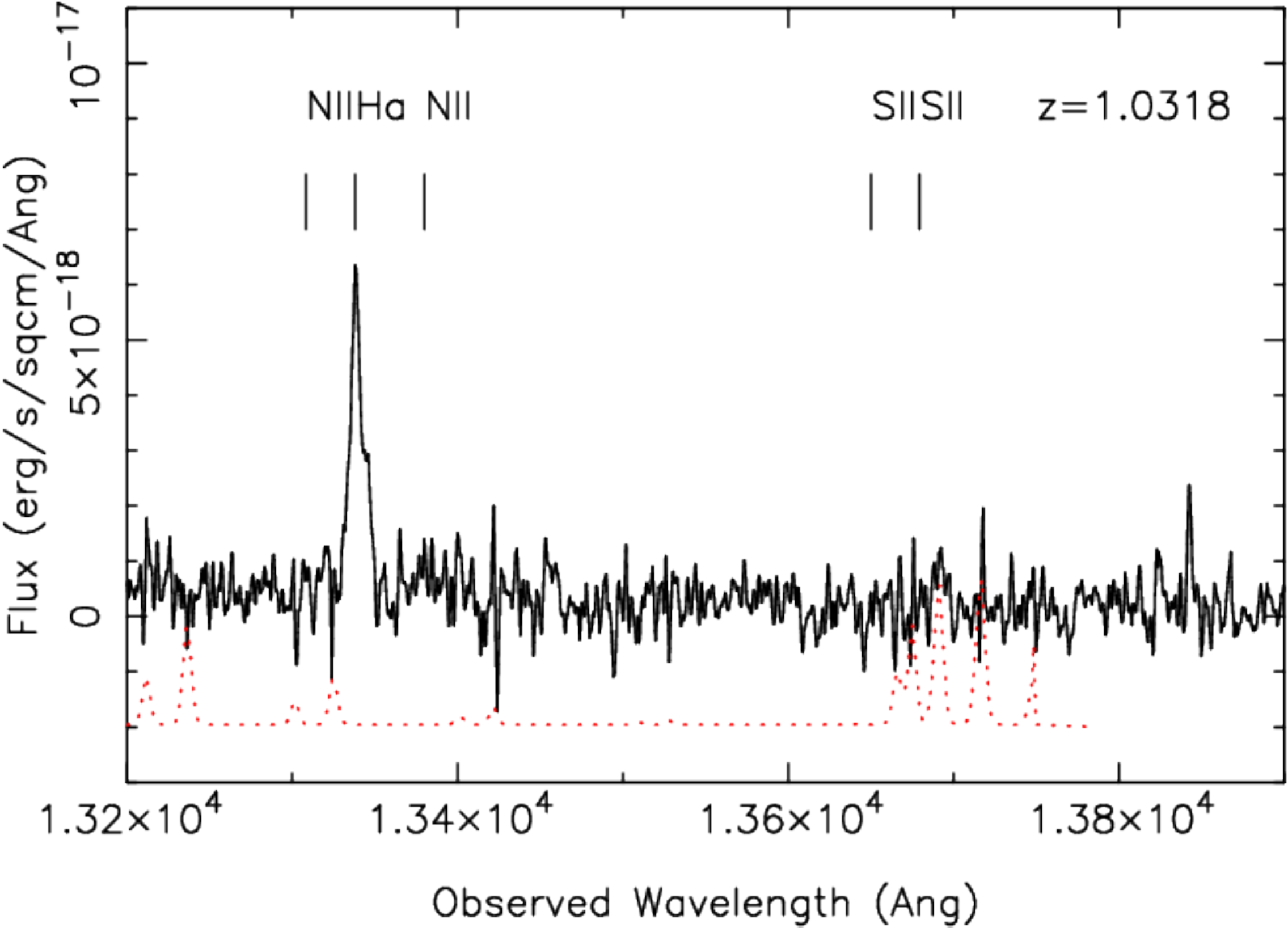}
\caption{{\bf \ha\ emission line detected in the X-Shooter spectrum of the DLA host towards Q2352$-$0028.} Few lines are detected in the X-Shooter spectrum of this absorbing-galaxy. The only reliable detections are from \ha\ and the \oii\ doublet. The other lines lead to upper limits.
}
\label{f:Q2352_1D_XSH}
\end{center}
\end{figure}

\subsubsection{Comparison of Phases} 

Abundances of galaxies are typically inferred from oxygen in nebular (i.e. emission line) spectra of \hii\ regions. It is important to note that \hii\ region may suffer from self-enrichment (Kunth \& Sargent 1986) and the bulk of metals in star forming galaxies may in fact be in the neutral gas (up to 90--95\%). It is therefore important to be able to compare these two metallicity indicators both in the local universe and at high-redshift. 

In the Milky Way, the oxygen abundances in Orion's \hii\ regions (Bautista \& Pradhan 1995) are in good agreement with those measured from UV absorption lines in the local interstellar medium (Moos et al. 2002). In the past years, IZw18 has been studied in great details (Aloisi et al. 2003; Lecavelier des Etangs et al. 2004; Lebouteiller et al. 2004) using FUSE observations. The neutral gas \hi\ is found to be already enriched in heavy elements, i.e. not primordial. The $\alpha$-elements traced by O, Ar and Si are $\sim$ 0.5 dex lower in the neutral interstellar medium (ISM) but the Fe abundance is found to be similar to the ionised phase, probably because Fe is release on a longer time-scale. A star formation at least $\sim$ 1 Gyr old is therefore required to explain the Fe abundance of the ISM. Lebouteiller et al. (2009) studied a sample of low metallicity galaxies and find that abundances in neutral gas are equal or lower than in the ionised gas and confirm a plateau below which metals are not found. We note that these results seem to put into question the reliability of \hii\ regions to be a good tracer of the ISM metallicity, but the difference in abundance between the neutral and ionised gas may be a by-product of the sample selection. Other studies at z=0 are underway (James et al. 2011) for 9 galaxies based on ACS and COS observations\footnote{see https://webcast.stsci.edu/webcast/detail.xhtml?talkid=3146 \&parent=1}. The authors find that overall the \hi\ abundances are similar to those in \hii\ region except for N, higher in \hii\ regions, (suggesting enrichment in \hii\ regions by Wolf-Rayet Stars). The oxygen metallicity in \hi\ and \hii\ agree well. Early results indicate that the difference between distinct gas phase are smaller towards the nucleus of the galaxies probably because there is more efficient mixing. These findings imply that we might indeed expect emission and absorption abundances to be in good agreement, at least on scales of a few kpc. Similarly, Schulte-Ladbeck et al. (2004, 2005) compare STIS abundances of the DLA towards the quasar HS 1543$+$5921 and the emission line abundances of the \hii\ region in the dwarf spiral SBS 1543$+$593. In that case, the angular separation is only $\sim$3kpc. The authors find that the neutral and ionised gas abundances agree, maybe because of more efficient mixing from rotation in that spiral. Bowen et al. (2005) utilised all the HST/STIS data obtained for this object and were able to confirm these conclusions.

For the quasar absorption line galaxies in our sample, both \hii\ and \hi\ metallicities free from dust-bias are available. This allows us to compare the metallicity at the centre of the galaxy (seen in emission) with the metallicity at a radius of a few dozen of kpc corresponding to the impact parameter (seen in absorption). At present, there are only nine systems (including five from our recent publications) for which both these measurements are available. For the three of these (Q0302$-$223, Q0452$-$1640 and Q2352$-$0028), we have presented here robust \hii\ metallicity estimates based on several emission line indicators observed with X-Shooter and reported in Table~\ref{t:HII_Metals}. In all three cases, the metallicity derived for the \hi\ gas is slightly lower than the one measured in the ionised gas. If we use these data to derive abundance gradients over 25 kpc, 16 kpc and 12 kpc (see Table 1), we obtain $-$0.02$\pm$0.08, $-$0.03$\pm$0.08 and $<-$0.04 dex/kpc for Q0302$-$223, Q0452$-$1640 and Q2352$-$0028, respectively. These slopes depict flat gradients. 

In the three cases where \nii\ is detected in the SINFONI data, we can compare the internal gradients derived from 2D information within the galaxies in the section above with the gradients deduced from a comparison of the ionised and neutral phases. We recall that we derived $-$0.11$\pm$0.17, $<$$-$0.10 and $-$0.07$\pm$0.35 dex/kpc for the ionised gas abundance within the galaxies, while P\'eroux et al. (2012) reported $-$0.03$\pm$0.08, $+$0.01$\pm$0.80, and $<-$0.04 dex/kpc over 16 kpc, 39 kpc and 12 kpc between the \hi\ and \hii\ phases for Q0452$-$1640, Q1009$-$0026 and Q2352$-$0028, respectively. While these gradients are spanning different physical scales, they agree within the admittedly large error bars. This suggests that neutral gas abundances are good tracers of the
metallicity of a galaxy. Since the internal gradient within the ionised gas aligns well
with the gradient between the ionised gas in the nucleus to the neutral gas in
the halo, one can assume that the latter is an acceptable proxy of the former. However, this assumes that the chemical evolution of the neutral gas mirrors that of
the ionised gas, i.e. that a gradient in the neutral gas also exists.

\section{Dust}
\subsection{Element Ratio}

By analogy with studies of local galaxies,
we expect interstellar dust to be a pervasive component of
\hi\ gas of galaxies observed in absorption along the line-of-sight to the background quasar. The amount of dust is empirically inferred from the relative abundances of iron-peak elements with respect to zinc, which is not depleted much on to dust grains in the interstellar medium of the Milky Way (Meyer \& Roth 1990). 

For the DLA towards Q0302$-$223, the elemental ratios [Fe/Zn]=$-$0.64$\pm$0.17, [Cr/Zn]=$-$0.40$\pm$0.17 and [Mn/Zn]=$-$0.76$\pm$0.17 indicate a rather large amount of depletion in the system with respect to the mean values for DLAs (Noterdaeme et al. 2008). Pettini et al. (2000) note a strong difference in the dust content of the two main components in this system: the component at v=0 km/s has [Fe/Zn]=$-$0.75, [Cr/Zn]=$-$0.52 and [Mn/Zn]=$-$0.84 while the one at v=$-$36 km/s has [Fe/Zn]=$-$0.25, [Cr/Zn]=$-$0.01 and [Mn/Zn]=$-$0.43. The authors recall that in the local interstellar medium the so-called Routly-Spitzer effect (Routly \& Spitzer 1952) also indicates reduced depletions in high-velocity interstellar clouds. For the DLA towards Q0452$-$1640, the elemental ratios [Fe/Zn]=$-$0.38$\pm$0.11, [Cr/Zn]=$-$0.18$\pm$0.12 and [Mn/Zn]=$-$0.58$\pm$0.14 indicate a rather modest amount of depletion. For the sub-DLA with log \nhi=19.81$^{+0.14}_{-0.11}$ towards Q2352$-$002, the elemental ratios [Fe/Zn]$>+$0.14 and [Cr/Zn]$>+$0.02 indicate a rather small amount of depletion, atypical of the lower \nhi\ systems (Meiring et al. 2009).

\subsection{Balmer Decrement}

The X-Shooter spectra presented here allow for a detailed study of the dust content of the \hii\ regions in the absorbing-galaxies. Extinction in
z$\ge$2 star-forming galaxies is commonly estimated from rest-frame UV colors and an application of the Calzetti et al. (2000)
starburst attenuation law. The Calzetti law appears to provide
a fairly accurate description on average of the reddening and
attenuation of the UV stellar continuum in both nearby and
distant starburst galaxies (Reddy \& Steidel 2004; Reddy et al.
2006). The degree of dust extinction in star-forming regions
can also be estimated from the Balmer lines of hydrogen,
because these strong optical lines have intrinsic ratios that are
well described by atomic theory. Under the assumption of Case
B recombination (Osterbrock 1989) and for T = 10, 000 K,
the Balmer ratios are set. Any deviations in the observed line
ratios are then attributed to dust extinction. Calzetti (2001)
demonstrates that, in local star-forming galaxies, the stellar
continuum suffers less reddening than the ionised gas, expressed
as $E(B-V)_{\rm star} = 0.44E(B-V)_{\rm gas}$. Furthermore, Calzetti
(2001) suggests that the reddening of the Balmer lines in nearby
UV-selected starbursts is better described by a foreground dust
distribution and traditional Milky Way extinction curve (Cardelli, Clayton \& Mathis 1989). On the other hand, Erb et al. (2006b) present
evidence that, in z$\sim$2 star-forming galaxies, $E(B-V)_{\rm star} \sim E(B-V)_{\rm gas}$ and a Calzetti et al. starburst extinction law applied
to both UV-continuum and \ha\ emission lines gives rise to the
best agreement between UV- and \ha-derived SFRs. 

In the case of quasar absorbers, York et al. (2006) have used a sample of more than 800 \mgii\ absorption systems with 1$<$z$_{\rm abs}$$<$2 in the Sloan Digital Sky Survey (SDSS) to show that the average extinction curve in these systems is similar to the SMC extinction curve, with no 2175 \AA\ feature and a rising UV extinction curve below 2200 \AA. We note, however, that a small number of absorbers have been reported to have the 2175 \AA\ features (e.g. Junkkarinen et al. 2004; Wang et al. 2004; Srianand et al. 2008 and Kulkarni et al. 2011). For larger \nhi\ systems traced by DLAs, the evidence for extinction are weak (Frank \& P\'eroux 2010; Khare et al. 2012), which is possibly a limitation intrinsic to the SDSS survey and quasar-selection itself (Noterdaeme et al. 2009).

Here, we use the detection of both \ha\ and \hb\ in the absorbing-galaxy to compute the Balmer decrement, which can allow us to infer the dust extinction by comparing Balmer line ratios to their predicted values via atomic physics. The colour excess can be computed from the following equation:

\begin{equation}
E(B -V)=\frac{1.086}{k (\lambda_{H\beta})-k(\lambda_{H\alpha})} {\rm ln} (\frac{H\alpha}{2.88 \times H\beta})
\end{equation}

These values are the intrinsic \ha/\hb, and k($\lambda$) is derived assuming an extinction curve of the SMC at wavelength $\lambda$. We derive the parameters k ($\lambda_{H-\alpha}$) and k ($\lambda_{H-\beta}$)  from an interpolation of the empirical extinction curves of Pei (1992) and the relation k($\lambda$) = A($\lambda$)/$E(B-V)$ from Calzetti (2004) as described in Gharanfoli et al. (2007). We estimate the extinction correction assuming a Milky Way type (MW-type) law. The resulting gas-phase colour excesses are estimated to be $E(B-V)$=0.1$\pm$0.8, 0.8$\pm$0.4 and $>-$0.4 for the absorbing-galaxies towards Q0302$-$223, Q0452$-$1640 and Q2352$-$0028, respectively. These values are listed in the last column of Table~\ref{t:HII_Metals}. These estimates have an opposite trend to the values derived from abundance ratios in the neutral gas phase in the section above. Indeed, the system towards Q0302$-$2238 presents a higher level of extinction in absorption but a lower value of $E(B-V)$, while the absorbing-galaxy towards Q0452$-$1640 indicates a more modest amount of extinction in absorption but a higher value of $E(B-V)$. Thus, the gas directly in front of the quasar appears to have different dust properties than the absorbing-galaxy's gas probed at the line of sight in our emission-line spectroscopy.
In the case of the object towards Q2352$-$0028, the $E(B-V)$ derived is minimally constrained due to the absence of detection of the \hb\ emission line.

%%%%%%%%%%%%%%%%%%%%%%%%
%%%%%%%%%%%%%%%%%%%%%%%%
\subsection{Quasar Reddening}

\begin{figure*}
\begin{center}
\includegraphics[height=17.5cm, width=7.5cm, angle=-90]{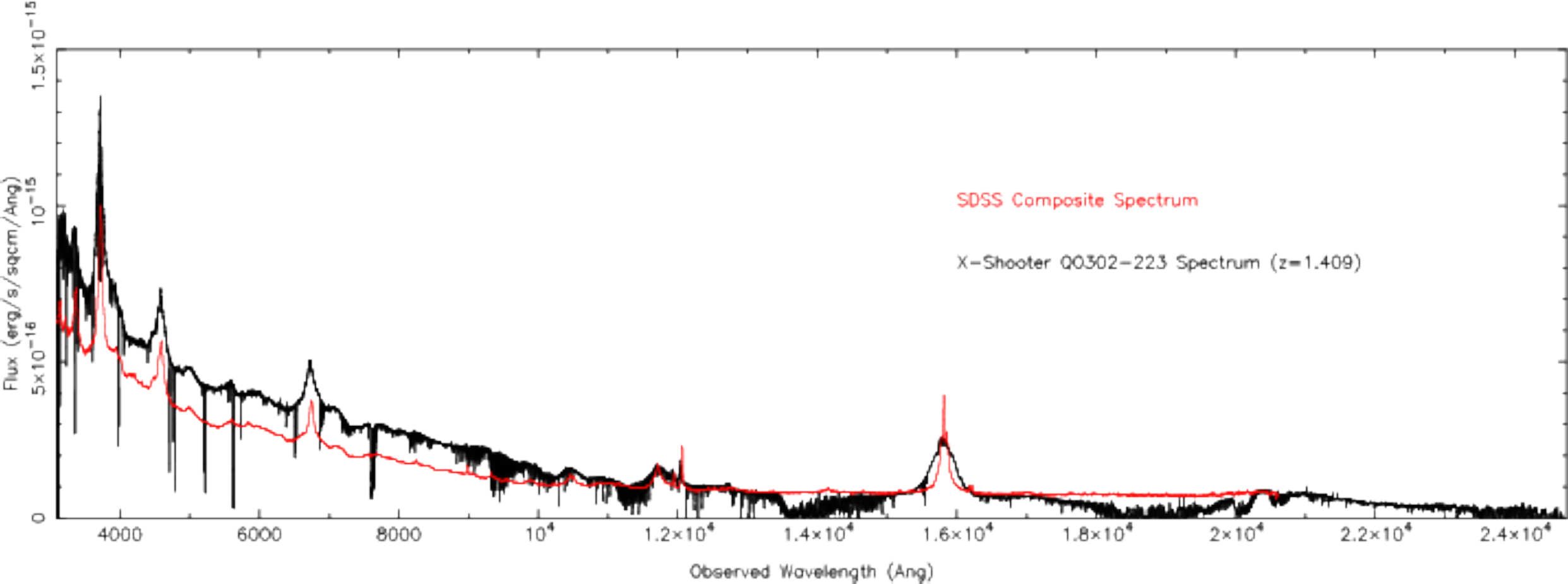}
\includegraphics[height=17.5cm, width=7.5cm, angle=-90]{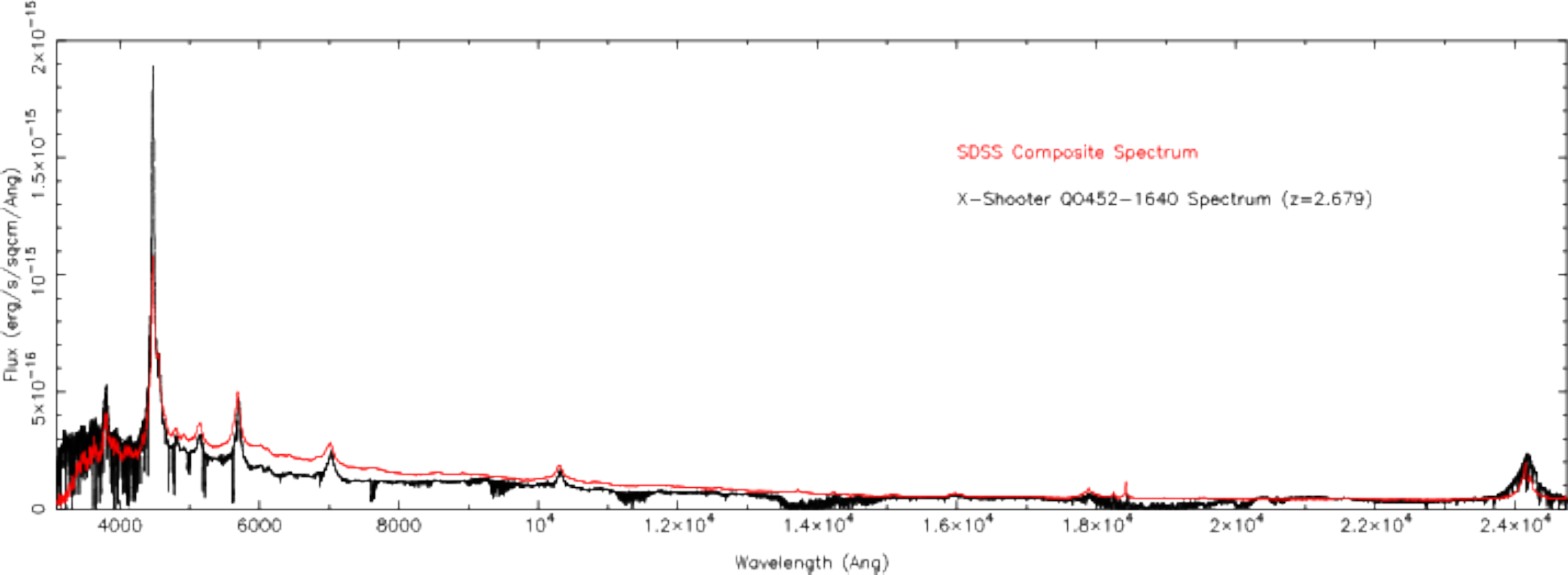}
\includegraphics[height=17.5cm, width=7.5cm, angle=-90]{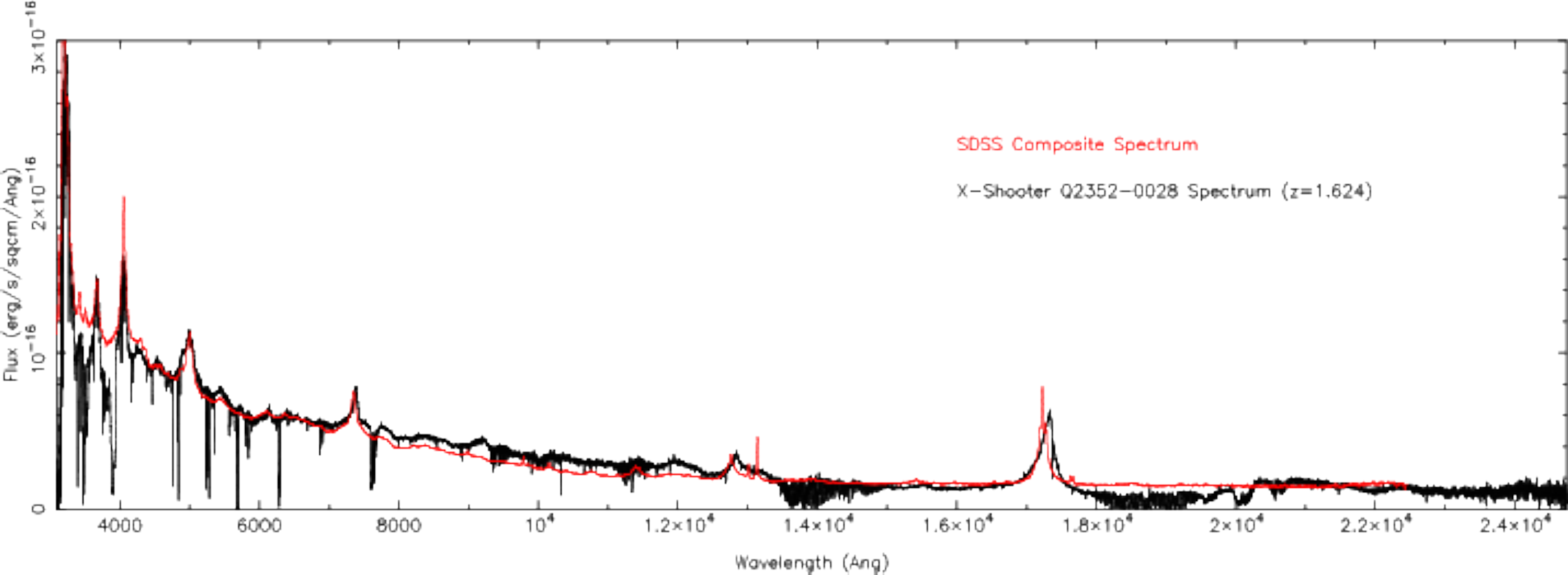}
\caption{{\bf The full X-Shooter spectra of the three targeted quasars.} The spectra are smoothed with a 3-pixel boxcar for display purpose. Overplotted on each panel is the SDSS quasar composite spectrum of Vanden Berk et al. (2001) redshifted to the object position and arbitrarily scaled to the observed fluxes. The emission redshifts from the quasars are indicated in the labels of each panel.}
\label{f:whole_spec}
\end{center}
\end{figure*}

Figure~\ref{f:whole_spec} presents the full X-Shooter quasar spectra flux-calibrated in erg/s/cm$^2$/\AA. Overplotted on each panel is the SDSS quasar composite spectrum of Vanden Berk et al. (2001) redshifted to the object position and arbitrarily scaled to the observed fluxes. Because of the large wavelength coverage of the X-Shooter observations, the spectra offer a strong constraint on the shape of the quasar continuum. Indeed, it is a common practice to compare the shape of an individual quasar spectrum with a composite to derive the reddening of the quasar (Fynbo et al. 2010; Noterdaeme et al. 2012; Fynbo et al. 2011). This information is difficult to relate to the integrated effect of dust due the intervening absorbers, because it cannot be differentiated from the intrinsic colour of the quasar which is known to vary from one object to another. A more robust approach to an estimate of the reddening from the measure of the quasar continuum slopes has to come from a statistical approach to overcome the inherent difference from one object to another (Murphy \& Liske 2004; York et al. 2006;  Frank \& P\'eroux 2010; Khare et al. 2012; Kulkarni et al. 2011).

\section{Conclusion}

We have presented X-Shooter emission-line spectra covering the observed wavelength 300 nm to 2.5 $\mu$m for three absorbing-galaxies reported by P\'eroux et al. (2011a, 2012). The slit was oriented to obtain a spectrum of both the background quasar and the galaxy responsible for the high-\nhi\ quasar absorber. These data thus allow for a robust estimate of \hii\ abundance in these and allow for a comparison of the metallicities in the neutral and ionised phase of the same objects. Our results suggest that the abundances derived in absorption along the line-of-sight to background quasars are a reliable measures of the overall galaxy metallicities. The 2D metallicity maps based on SINFONI observations previously published (P\'eroux et al. 2011a, 2013a), show small negative metallicity gradients. The flat slopes are in line with the differences observed between the two phases of the gas. These results suggest that a comparison of the \hi\  and \hii\ metallicities is a robust indicator of the internal gradients. Finally, we use several indicators to measure the quantity of dust in the ionised and neutral phases of these systems. The presence of dust in the \hi\ phase can be estimated from the observed depletion of refractory elements. We find measures from the Balmer decrement of the galaxies to be in line with values derived from element ratios in the \hi\ gas. We note however, that the smaller \nhi\ column density towards Q2352$-$0028 show little indication of dust depletion, contrary to what is expected from such sub-DLAs.

\section*{Acknowledgements}
We are grateful to Nicolas Bouch\'e for his contribution to the overall project. We would like to thank the Paranal and Garching staff at ESO for performing the observations in service mode and the instrument
team for making a reliable instrument. VPK acknowledges partial support from the U.S. National Science Foundation grants AST/0908890, AST/1108830 (PI: Kulkarni). This work has benefited from support of the 'Agence Nationale de la Recherche' with reference ANR-08-BLAN-0316-01.

\bsp

\label{lastpage}
\end{document}